\documentclass[a4paper,11pt]{article}
\pdfoutput=1 

\usepackage{jcappub} 

\usepackage[T1]{fontenc} 

\usepackage{xcolor}
\usepackage{wrapfig}
\usepackage{multirow}
\usepackage{caption}
\usepackage{subcaption}
\usepackage{sidecap}
\usepackage{bm}
\usepackage[utf8]{inputenc}
\usepackage[normalem]{ulem}
\usepackage{floatrow}
\usepackage[inline]{enumitem}  
\usepackage{multicol}

\usepackage{mathrsfs}
\graphicspath{ {images/} }

\usepackage{todonotes}
\usepackage{appendix}
 
\def\x{{\bf x}}

\title{Reconstruction of Dark Matter and Baryon Density From Galaxies: A Comparison of Linear, Halo Model and Machine Learning-Based Methods}

\author[b,c]{Jordan Krywonos}
\author[a]{Yurii Kvasiuk}
\author[b,c]{Matthew C. Johnson}
\author[a,d]{Moritz M\"unchmeyer}

\affiliation[a]{Department of Physics, University of Wisconsin-Madison, Madison, WI 53706, USA}
\affiliation[b]{Perimeter Institute for Theoretical Physics, Waterloo, ON N2L 2Y5, Canada}
\affiliation[c]{Department of Physics and Astronomy, York University, Toronto, ON M3J 1P3, Canada}
\affiliation[d]{NSF-Simons AI Institute for the Sky (SkAI), 172 E. Chestnut St., Chicago, IL 60611, USA}

\emailAdd{jkrywonos@perimeterinstitute.ca}
\emailAdd{kvasiuk@wisc.edu}
\emailAdd{mjohnson@perimeterinstitute.ca}
\emailAdd{muenchmeyer@wisc.edu}

\abstract{For many analyses in cosmology it is necessary to reconstruct the likely distribution of unobserved fields, such as dark matter or non-luminous baryons, from observed luminous tracers. The dominant approach in cosmology has been to use the so-called halo model, which assumes radially symmetric profiles centered around luminous tracers such as galaxies. More recently, field-level machine learning methods have been proposed that can learn to estimate the unobserved field after being trained on simulations. However, it is unclear whether machine learning methods indeed significantly improve over linear methods or the halo model. In this paper we make a systematic comparison of different approaches to reconstruct dark matter and non-luminous baryons, from galaxy data using the CAMELS simulations. These simulations are in a $25\ \texttt{Mpc/h}$ box, allowing us to compare performance on the mildly non-linear scales $(k\sim 0.4\ \mathrm{h/Mpc})$ down to the size of individual halos. We find the best results using a combined GNN-CNN approach. We also provide a general analysis and visualization of the relationship of matter, non-luminous baryons, halos, and galaxies in these simulations to interpret our results.}

\begin{document}
\maketitle
\flushbottom

\section{Introduction}

Within the $\Lambda$CDM paradigm, dark matter provides the scaffolding on which structure forms, while baryonic matter, which we will refer to as `gas', provides the medium from which astrophysical objects like stars and galaxies form. Dark matter, and the majority of gas, is not directly observable. Instead, we observe luminous tracers, e.g.  starlight from galaxies or thermal radiation from hot gas, that are embedded within the underlying non-luminous dark matter and gas. In this paper we study how luminous tracers can be used to make field-level inferences about the distribution of dark matter and gas, using both classical and machine learning methods. 
Making field-level inferences requires a model for the connection between luminous observable tracers and the underlying unobservable dark matter and gas fields provided by simulations. These models can be analytical, empirical, or numerical. In all cases, the observable input is a galaxy catalog (real or simulated), and the output is an estimate of the continuous dark matter or gas field. 

At a basic level, the problem we wish to solve is to find the statistically ideal non-linear map between a point cloud (galaxies) and a pixelated continuous field (dark matter or gas). While we will also evaluate traditional halo and transfer function based approaches, our focus will be on machine learning. A class of machine learning (ML) architectures well-suited to point clouds are graph neural networks (GNN); see e.g.~\cite{battaglia2018relationalinductivebiasesdeep}. The GNN architecture encodes galaxy features in nodes, and relational information (e.g. relative distance) in vertices connecting them. GNNs have successfully been used to infer the mass of individual halos from galaxy features e.g.~\cite{garuda2024estimatingdarkmatterhalo,VD_hm} or cosmological parameters e.g.~\cite{VD_siom}. A class of ML architectures well-suited to pixelated continuous fields, a type of regular grid-structured data, are convolutional neural networks (CNN). CNNs have been applied to a wide variety of problems in cosmology e.g. Refs.~\cite{ravanbakhsh2017estimatingcosmologicalparametersdark,schmelzle2017cosmologicalmodeldiscriminationdeep,Gupta_2018,Ribli_2019,Fluri_2019,Matilla_2020,villaescusanavarro2021multifieldcosmologyartificialintelligence,villaescusanavarro2021robustmarginalizationbaryoniceffects,Villanueva_Domingo_2021,Lazanu_2021,Lu_2022}. In Ref.~\cite{Kvasiuk:2024kwe}, we introduced a hybrid GNN-CNN architecture that maps a galaxy catalog directly to the dark matter field in three dimensions. Training the network on the CAMELS suite of cosmological simulations~\cite{CAMELS_presentation,CAMELS_DR1,CAMELS_DR2,CMD}, we demonstrated that such a GNN-CNN hybrid architecture could outperform the linear transfer function or a gridded halo mass assignment.

In this paper, we undertake a detailed study of the capabilities of a GNN-CNN hybrid architecture to reconstruct the dark matter field from galaxies on the scales of 25 \texttt{Mpc/h}, down to the scale of individual halo, comparing the fidelity of the reconstruction to a variety of other techniques based on the halo model. In Section~(\ref{sec:overview}), we provide an overview of techniques from the literature to reconstruct unobserved fields. Throughout this paper, we use simulations for the galaxies and dark matter fields so we can benchmark against a ground truth dark matter, these simulations are described in Section~(\ref{sec:simdata}). The ML and traditional methods we use to reconstruct the unobserved fields are outlined in Section~(\ref{sec:methods}). In the results, Section~(\ref{sec:DMrec}), we provide a detailed visualization of the reconstructed dark matter field using different techniques, highlighting the features that are captured and those that are not. We demonstrate that the GNN-CNN reconstruction trained on CAMELS is superior to halo-model based approaches. This is also shown for the reconstructed gas field in Section~(\ref{sec:gas}) and we investigate the GNN-CNN's robustness with respect to unknown feedback parameters in Section~(\ref{sec:robust}). We comment on cosmological and astrophysical parameter inference using this reconstruction in Section~(\ref{sec:paramdepinf}). In Appendix~(\ref{app:bv}), we also discuss explicitly how baryonic uncertainty can be taken into account by marginalizing over a bias parameter, in the specific case of kinetic Sunyaev-Zeldovich velocity reconstruction, which is a major physical application for our method. Finally, we summarize our results in Section~(\ref{sec:conc}) and discuss future avenues to apply this to real data.

\section{Overview of Approaches to Reconstruct Unobserved Fields}
\label{sec:overview}

The goal of this paper is to reconstruct the unobserved matter distribution $\delta_m$ and the unobserved baryon distribution $\delta_e$ from the observed galaxy density $\delta_g$. We assess the performance of different reconstruction methods using simulations, with the future goal to apply the best method to galaxy observations. In this section, we first give a high-level overview of several possible reconstruction methods.

\subsection{Linear Transfer Function}

A common approach is to use a linear transfer function in Fourier space to approximate the target field from the observed field:
\begin{equation}
\hat{\delta}_{\mathrm{m}}(\mathbf{k}) = T(k) \, \delta_{g}^{\rm obs}(\mathbf{k})
\end{equation}
The choice of linear transfer function that minimizes $\langle|\hat{\delta}_{\mathrm{m}}(\mathbf{k}) - \delta_{\mathrm{m}}(\mathbf{k}) |^2\rangle$ is the ratio of cross- to auto-power spectra
\begin{equation}\label{eq:transferk}
    T(k) = \frac{P_{gm}(k)}{P_{gg}(k)}.
\end{equation}
The transfer function in Eq.~(\ref{eq:transferk}) can be derived by assuming that $g$ and $m$ are correlated Gaussian random fields and that $g$ is observed with some shot noise (included in $P_{gg}$). A different possibility would to use the transfer function of form
\begin{equation}
    T(k) = \frac{\sqrt{P_{mm}(k)}}{\sqrt{P_{gg}(k)}}.
\end{equation}
as studied for example in \cite{Sharma:2024kwj} in the context of mapping N-body to hydrodynamic simulations, which would make the power spectrum correct by definition. A further recent example of the use of a transfer function to model the gas distribution can be found in Ref.~\cite{Liu:2025gba}. The transfer function can either be fit directly to simulations or calculated using the halo model - which is introduced in the following Section (\ref{sec:halomodel}).

Of course, a linear transfer function does not improve the cross-correlation coefficient between the observed field and the target field, which is invariant under this re-scaling. We also note that a linear transfer function can always be added to other methods, to enforce that the reconstructed field has any desired power spectrum. Note however, that one does not always want the power spectrum of the reconstructed field to be equal to the target field. Often, the output should be inverse covariance weighted so that noisy (small-scale) modes are suppressed. In the above transfer function Eq. \eqref{eq:transferk}, this will happen if $P_{gg}$ includes the galaxy shot noise. 

\subsection{Dark Matter and Gas Reconstruction Using Halo Models}\label{sec:halomodel}

A dominant paradigm for understanding the connection between luminous tracers and the dark matter and gas distribution is the halo model of large-scale structure (see e.g. Refs.~\cite{Cooray:2002dia,Asgari_2023}). Within this analytic construction, all dark matter is assumed to be bound into halos whose positions are correlated by the long-wavelength  density field. Each halo is populated with gas or galaxies according to a set of theoretical prescriptions, with parameters calibrated by matching a variety of observations. For galaxies this is referred to as the `Galaxy-Halo Connection' (see e.g. Ref.~\cite{Wechsler_2018} for a review) and the (probabilistic) set of rules for populating halos with galaxies is the Halo Occupation Distribution (HOD). The HOD depends on the properties of the galaxy sample and dark matter mass of the halo, but might also depend on more complex physics, e.g. the assembly history (a recent assessment can be found here~\cite{Zhai:2024knl}). For gas, a simple model could be a parameterized spherically-symmetric density profile that depends on halo mass (a recent example can be found in Ref.~\cite{oppenheimer2025introducingdescriptiveparametricmodel}). This profile may be derived based on physical principles, e.g. hydrostatic equilibrium, or measured from hydrodynamic simulations or observations. For computing summary statistics, the halo model provides a physically-motivated description that can be fit to data in order to make inferences about cosmology, galaxies, and gas. However, there are limitations to this approach such as the non-conservation of mass and a poor description of intermediate scales (the halo model can be modified to mitigate these shortcomings, e.g.~\cite{Ginzburg_2017,Chen:2019wik}).

\paragraph{Halo painting.} 

A method to make a template for dark matter or baryons from galaxies is to identify the dark matter halo that corresponds to a group of galaxies, and then assign a NFW profile (for dark matter) or a gas profile (for baryons) centered on that halo. This mirrors how the cross-correlation of galaxies and dark matter or baryons is calculated in the halo model. This type of `painting' on the location of halos is commonly employed in creating mock observables from dark matter-only simulations, see e.g.~\cite{Sehgal_2010,Stein_2018,Stein_2020,Yasini2020}. In pixel space, we can think of the linear transfer function defined in Eq.~\eqref{eq:transferk} as a simple example of this, where an identical density profile is painted on the position of each galaxy. This can be seen by writing the galaxy field as a sum over delta functions
\begin{equation}
    \delta_g^{\rm obs} (\vec{x}) = \sum_{i=1}^{N_g} \delta^{(3)} (\vec{x} - \vec{x}_i)
\end{equation}
and noting that the Fourier transform of the transfer function is spherically symmetric and given by
\begin{equation}
    T(|\vec{x}|) \equiv \int \frac{d^3k}{(2\pi)^3} \frac{P_{gm}(k)}{P_{gg}(k)} e^{i \vec{k}\cdot\vec{x}}
\end{equation}
The dark matter density field is then
\begin{eqnarray}
    \delta_m(\vec{x}) &=& \int d^3y \ T(|\vec{x}-\vec{y}|) \ \sum_{i=1}^{N_g} \delta^{(3)} (\vec{y} - \vec{x}_i) \\
    &=& \sum_{i=1}^{N_g} T(|\vec{x}-\vec{x}_i|)
\end{eqnarray}
The profile painted on top of each galaxy position is the inverse Fourier transform of the transfer function.

One can improve over this simple model by considering individual halos (with individual mass and/or concentration), inferred from a learned model of the galaxy-halo connection, to assign dark matter density profiles based on observed galaxy properties. This results in a \emph{non-linear} reconstruction method and is thus not captured by a transfer function. We explore this possibility by training a GNN model to infer the mass of NFW halos from the point cloud of galaxies, and then applying NFW profiles about the center of each inferred halo location. We compare this to an `exact' version of this technique, where the  density field is obtained by painting NFW profiles with parameters drawn from the true halo catalog of the simulation. We find that inferred halo masses somewhat degrade the reconstruction over assuming true (unobservable) halo masses. 

A final comment regarding halo painting is that there is implicit dependence at the field level on how one assigns mass and the corresponding truncation radius of dark matter halos (e.g. at the virial radius, a density-dependent radius, etc); a related ambiguity is the total fraction of dark matter mass in halos that also host galaxies versus those that do not. 

\subsection{Dark Matter and Gas Reconstruction Using Machine Learning}

Another set of approaches relies on ML methods. These are trained on simulations and can then be applied to galaxy observations. ML offers a powerful framework for handling high-dimensional problems and capturing complex relationships across a range of scales. For example, Refs.~\cite{InferringGalDH, Villanueva-Domingo_2022, Calderon:2019, 2023MNRAS.525.6015D, Hahn_2024, 2024A&A...686A..80W} use ML to improve on halo abundance matching by predicting halo masses from a variety of galaxy properties in addition to stellar mass. At the field level, Refs.~\cite{ono2024debiasingdiffusionprobabilisticreconstruction, park2023probabilisticreconstructiondarkmatter} employ diffusion models to reconstruct 2D dark matter fields from galaxy distributions, and Refs.~\cite{Wang:2023hgm,Shi:2025zoz} use CNNs to reconstruct dark matter density, velocity, and tidal fields from halo/galaxy samples.  

Galaxy catalogs are better represented on a graph than a grid - to capture their discreteness and the tabular nature of their observed properties. This structure is well captured by GNNs. In contrast, the target dark matter and gas fields are continuous fields defined on regular grids, which makes CNNs an appropriate choice. Therefore, a suitable framework for field level reconstruction is to combine the above ML methods and have a hybrid GNN-CNN model~\cite{Kvasiuk:2024kwe}. In this paper, we further explore the performance of the combined GNN-CNN method from Ref.~\cite{Kvasiuk:2024kwe} to directly learn the non-linear galaxy point cloud-to-mass/baryon density field map. We train this architecture on the CAMELS suite of simulations, and compare the fidelity of the reconstruction to that obtained from the linear transfer function and halo painting methods described above. We demonstrate that this ML-based method yields the best reconstruction of the density fields among the methods we explore.

\subsection{Relation to Forward Modeling}

One numerical approach that has been extensively explored in the literature is to forward-model galaxy surveys, sampling Gaussian initial conditions, evolving them with N-body simulations, creating mock galaxy surveys and comparing the mocks with galaxy data to find the most likely underlying dark matter distribution~\cite{Jasche_2013,2013ApJ...779...15J,Jasche_2014,2016MNRAS.455.3169L}. This approach is based on causal inferences (because samples are generated by evolving initial conditions with physical equations of motion) rather than linear correlation, and the statistical distribution over initial conditions is Gaussian, allowing precisely defined error bars. 

Typically, forward modeling works by evolving  \emph{initial conditions} $\delta_{init}(\x)$ to a final matter field by a deterministic forward model $\mathcal{F}(\Lambda_{F})$ (such as a gravitational N-body simulation) which depends on unknown \emph{forward model parameters} $\Lambda_{F}$ (such as the matter density parameter $\Omega_M$):
\begin{align}
\label{eq:forward}
    \delta_m(\x,t) = \mathcal{F}(\delta_{init}(\x),\x,t,\Lambda_{F}).
\end{align}
The small-scale galaxy distribution is often modeled in a second stochastic step, which samples the observable galaxies from the deterministic matter distribution: 
\begin{align}
  \mathcal{P}_g(\delta_{g}^{\rm obs}(\x,t)| \delta_m(\x,t), \Lambda_g).
\end{align}
Here $\Lambda_g$ are the astrophysical parameters of the stochastic galaxy model. By assuming Gaussian initial conditions one can then sample the posterior
\begin{align}
    \mathcal{P}(\Lambda_{F},\Lambda_g,\delta_{init}|\delta_{g}^{\rm obs})
\end{align}
which implies a reconstruction of $\delta_m(\x,t)$ through Eq. \eqref{eq:forward}. The forward model for the galaxy distribution $\delta_m \rightarrow \delta_g$, whether stochastic or deterministic, is the inverse of our reconstruction which goes $\delta_g \rightarrow \delta_m$. There are many models for $\delta_m \rightarrow \delta_g$ in the literature, working at different scales, some examples including \cite{Modi:2018cfi,Horowitz:2022fvl}. 
However, forward modeling is more computationally expensive than the direct estimate we are making here, and it is not clear whether these more complicated methods outperform direct reconstruction. Further, forward modeling the matter distribution is only practical on sufficiently large-scales.

\section{Simulation Data}\label{sec:simdata}

\subsection{Dataset}\label{sec:dataset}

In this work, we use the CAMELS IllustrisTNG-CV simulation suite \cite{CAMELS_DR1,CAMELS_DR2,CAMELS_presentation}. From this, we consider redshift $z=0$ snapshots and the FOF-SUBFIND sub-halo/galaxy catalogs. The simulations evolve $256^3$ dark matter and $256^3$ gas particles in a $25\ \texttt{Mpc/h}$ box. This suite contains 27 simulations that differ by initial random seed, but have fixed cosmological parameters.

To make the galaxy catalogs closer to observable catalogs, we apply the selection criteria of Ref.~\cite{Wu_2024}:
\begin{itemize}
    \item The stellar and dark matter half-mass radii are greater than two Plummer radii ($R_p=0.74\ \texttt{kpc}$).
    \item There must be more than 200 stellar and dark matter particles.
    \item Assume each sub-halo is a separate galaxy.
\end{itemize}
This leaves, on average, 283 galaxies in a $25^3\ (\texttt{Mpc/h})^3$ volume. The galaxy dataset is not fully realistic, but can broadly be compared to data from ongoing and future high number density photometric surveys. For instance, the number density of galaxies in our dataset is around $10$ times larger than DESI at $z=1$, but below the expected value from Rubin Observatory \cite{M_nchmeyer_2019, Hadzhiyska_2023}. 

To compare the performance of ML or traditional methods to reconstruct the dark matter field, we need a true dark matter 3D grid for comparison. In hydrodynamical simulations, each particle represents the volume of the Voronnoi cell. So unlike in dark matter only N-body simulations, it does not make sense to use counts-in-cells (CIC) to assign the density to a grid. We instead use the CAMELS Multifield Dataset (CMD) \cite{CAMELS_presentation,CMD}, which approximates the Voronnoi cell by taking the radius to the 32nd nearest particle and assigning density to the grid based on the particles within that radius. We use the $256^3$ resolution. For these CMD grids, 3 of the 27 simulations do not match the galaxy catalogs, so we only use 24 simulations \footnote{We confirmed with the author of Ref.~\cite{CMD} that the 24 CMD simulations we use do correspond to the galaxy catalogs.}. To summarize, we use the following data: 

\paragraph{Input: Observed galaxy point cloud.} We define galaxies as sub-halos of the gas particles in the CAMELS simulation. A galaxy is located at the position of the center of the sub-halo. Additionally, each galaxy is associated with a stellar mass, given by the number of stellar particles in the sub-halo, among other properties including $r$, $g$, $z$ magnitudes, velocity dispersion, and stellar half-mass radius.

\paragraph{Output: Dark matter and baryon fields on a regular grid.} Our target fields will be the unobserved dark matter and baryon density, generated by assigning particles to a regular grid. The precise grid-assignment procedure influences the field distributions on small-scales. As discussed above, we do not use the common CIC procedure for grid assignment, but rather a Voronnoi volume inspired method, which is more appropriate for hydrodynamic simulations. In our analysis, this assignment is part of our definition of the `true' fields.

\subsection{Visualizing the Data}

\begin{figure}[h!]
\centering
  \includegraphics[width=\textwidth]{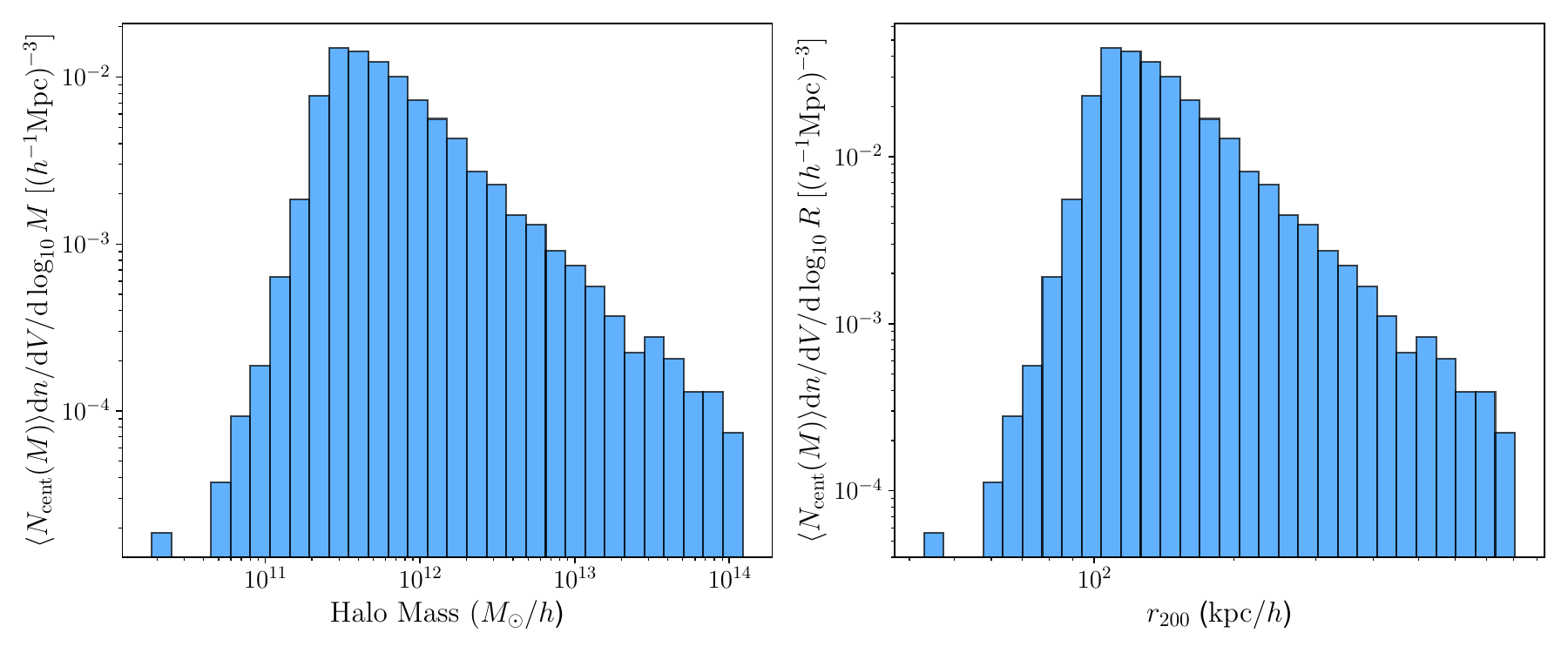}
  \caption{The probability of a halo hosting a central galaxy times the average number density distribution of halo masses (left) and $r_{200}$ radii (right).}
    \label{fig:Mh_R200_avgCVsims}
\end{figure}

To describe the halos in the simulations, we use the spherical overdensity halo mass $M_{200}$.
This is defined as the total mass enclosed within a sphere, where the average density is 200 times the critical density of the universe. Formally, it is given by $M_{\Delta}\equiv(4/3)\pi r_{\Delta}^3\rho_c\Delta$ where $\Delta=200$, $\rho_c(z) = 3H^2(z)/8\pi G$ is the critical density, and $H(z)$ is the Hubble parameter. In Fig.~(\ref{fig:Mh_R200_avgCVsims}) is the distribution of halo masses and $r_{200}$ radii, averaged over all 27 simulations, for halos that host a central galaxy. The halo masses are plotted as the number density per logarithmic mass so that they can be compared to the halo mass function. Note that the drop-off at small halo masses is due to our selection criteria of including only halos with a central galaxy, e.g. we show $\langle N_{\rm cent} (M) \rangle dn/dV/d\log_{10} M$ where $\langle N_{\rm cent} \rangle$ is the mean number of centrals as a function of halo mass.

Across all the CV simulations, with our selection cuts, the average fraction of dark matter mass inside halos is 79\% whereas for gas it is 36\%. For this calculation, we defined the halo by its $r_{200}$ radius. This demonstrates that the halo contribution is larger than the field contribution for dark matter, but the opposite applies for gas. The gas is much more dispersed outside of the halo. Therefore, for gas, it is more crucial to use field-based methods to construct the 3D density grid.

\begin{figure}[h!]
\centering
  \includegraphics[width=0.6\linewidth]{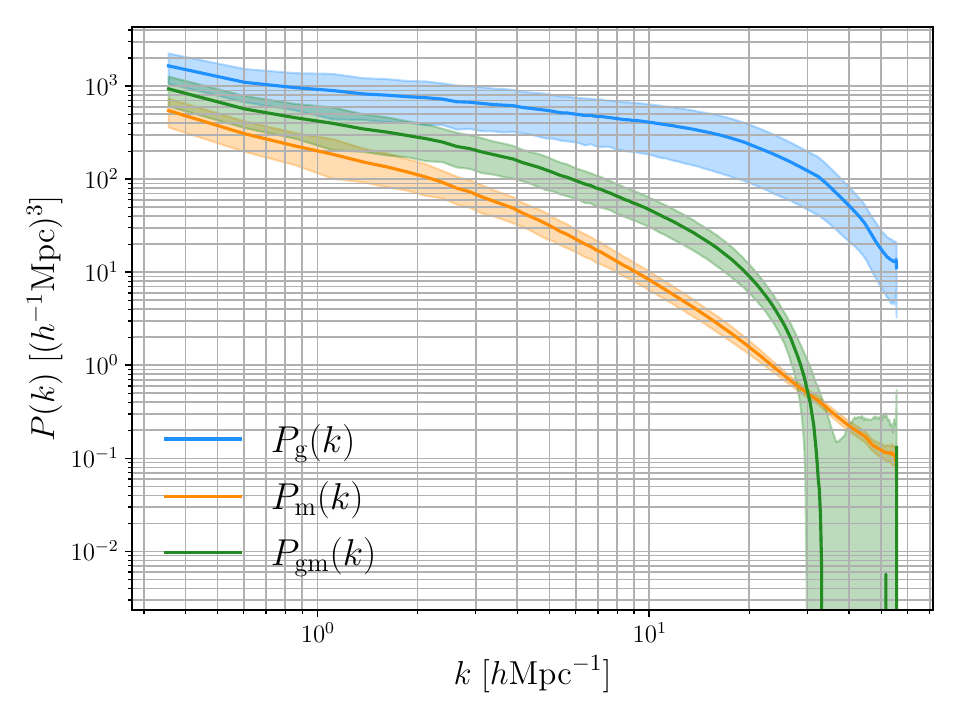}
  \caption{The auto-power spectra of the galaxy (blue) and dark matter (orange) fields. In green is their cross-correlation.}
  \label{fig:gal_dm_pspec}
\end{figure}

We aim to use galaxies to predict the underlying dark matter field. To get a better sense of the galaxy-dark matter connection in our simulation suite, we plot the cross-correlation power spectra in Fig.~(\ref{fig:gal_dm_pspec}). Here, the dark matter density field was constructed using CMD to grid the dark matter particles. The galaxy density field was obtained through CIC, by weighting galaxies by their stellar mass. On large-scales the $P_g(k)$ and $P_m(k)$  are related by linear galaxy bias, but the correlation decreases towards smaller scales which are influenced by baryonic effects and non-linear gravitational evolution.

\begin{figure}[h!]
\centering
  \includegraphics[width=0.9\linewidth]{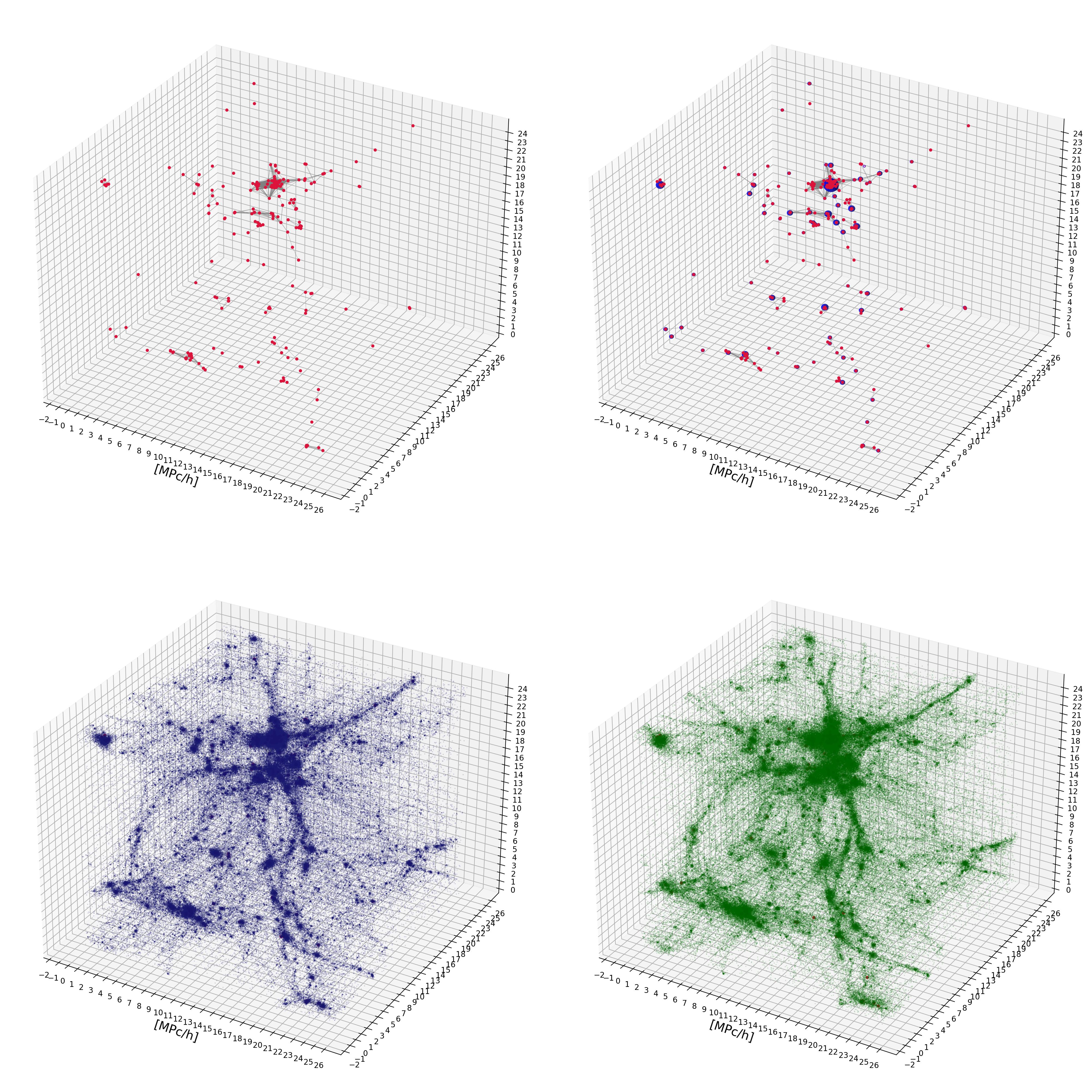}
  \caption{The distribution of galaxies (red points), dark matter halos (blue spheres), dark matter (blue) particles, and gas (green) particles in the simulation volume of $25^3\ (\texttt{Mpc/h})^3$. The radii of the spheres are proportional to $r_{200}$ of the halo.}
  \label{fig:snap3d}
\end{figure}

\begin{figure}[h!]
\centering
  \includegraphics[width=0.9\linewidth]{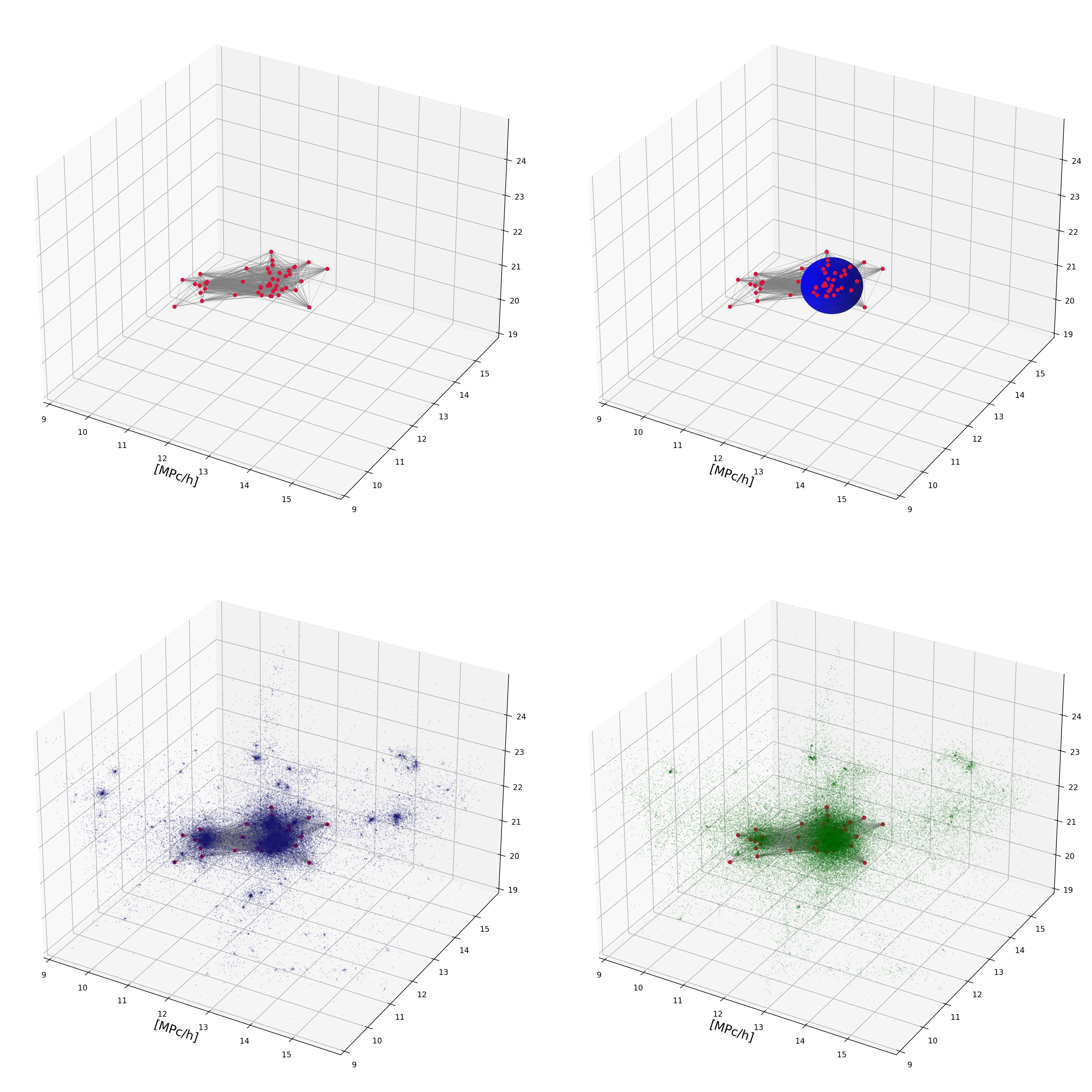}
  \caption{The distribution of galaxies (red dots) that belong to the largest halo in a simulation box. The halo is depicted as a blue sphere. Bottom left and right plots show the distribution of the dark matter and gas particles in the same zoomed-in region (blue and green correspondingly).}
  \label{fig:snap3d-zoom}
\end{figure}

To better understand the setup, it is useful to visualize the corresponding locations of the observed galaxy cloud, dark matter halos, and the dark matter and gas fields. Figs.~(\ref{fig:snap3d}) and (\ref{fig:snap3d-zoom}) depict the galaxy catalog after the selection cuts have been applied (red dots), the corresponding dark matter halos (depicted as spheres of $r_{200}$ radius), and the dark matter and gas densities. Fig.~(\ref{fig:snap3d}) shows the $(25\ {\texttt{Mpc/h}})^3$ volume, and Fig.~(\ref{fig:snap3d-zoom}) illustrates the zoomed-in region around the largest halo. To visualize dark matter and gas densities, we plot the locations of each 50th particle of the corresponding type. We can make a few useful observations. The locations of the galaxies provide a better guess of the underlying density field distribution than just dark matter halos. We also notice that the gas distribution is spatially more spread out, while the dark matter is more clumped.

\section{Methods}\label{sec:methods}

We consider different ML and non-ML methods to construct the continuous dark matter density field from a galaxy catalog. 
The non-ML benchmarks include using the CIC stellar mass-weighted galaxy field as a biased tracer of the dark matter field, constructing the dark matter field by applying a linear transfer function to the galaxy field, and painting the halo's NFW profile onto a grid.

The first ML approach uses a GNN to predict halo mass from galaxy input features, which allows us to paint on the NFW profiles while accounting for the fact that halo mass is not directly observable. The next method, called GNN-CNN, treats the galaxies as a point cloud, assigning them to graphs which encode the information into a proto-density field, which is then passed through a CNN. These approaches are outlined below. By comparing with the linear baseline, we will examine whether non-linear methods have an advantage.

\subsection{Performance Metric}

To compare the performance of different methods, the most common metric is the \emph{cross-correlation coefficient} of the reconstructed field $\hat{\delta}$ and the true field $\delta$. In terms of the power spectra it is given by
\begin{equation}
\label{eq:crosscorr}
    r^2(k)=\frac{P^2_{\rm true,est}(k)}{P_{\rm true}(k)P_{\rm est}(k)}
\end{equation}
where $P_{\rm true}(k)$ is the power spectrum of the field, $P_{\rm est}(k)$ is the power spectrum of the estimate of that field, and $P_{\rm true,est}(k)$ is a cross-spectrum between the two. The cross-correlation coefficient is the relevant performance metric if the goal is to increase the signal-to-noise ratio of a cross-correlation-based data analysis. We compute the power spectra using \texttt{Pylians} \cite{Pylians}. We do not deconvolve the window functions when plotting the power spectra, since they cannot be removed for the CMD fields. This means that there is a suppression of power at small-scales in the spectra plots.

\subsection{Baseline: Simple Grid Assignment and Linear Transfer Function} 
\label{sec_smw_and_linear}

As our baseline method, we use the mass-weighted tracer density. We thus assign the (stellar) mass-weighted galaxy point cloud to a regular grid, using the CIC procedure. We call the resulting field $\delta_{g}(\mathbf{k})$. We chose to mass-weight galaxies in the baseline because the shot noise would be quite large in the small Quijote volume otherwise, even though galaxy masses are only approximately observable. 

The second baseline approach is using a linear transfer function. This is done by applying a linear kernel to the mass-weighted galaxy overdensity field ($\delta_g(\vec{x})$) in $k$-space: 
\begin{equation}\label{eq: filter}
    \delta_m(\vec{k}) = \frac{P_{gm}(k)}{P_{gg}(k)} \delta_g(\vec{k}),
\end{equation}
then Fourier transforming back to real space to get the dark matter overdensity field ($\delta_m(\vec{x})$). The $P_{gm}(k)$ and $P_{gg}(k)$ are calculated using the CIC mass-weighted galaxy field and the CMD true dark matter field. 
The linear transfer function can give a dark matter density field whose power spectrum is a good approximation to the true one, but the field-level residuals when compared with the true underlying dark matter field can be large.

The CIC and linear transfer function methods will have the same cross-correlation coefficient, because they both are using the galaxy field as a tracer of the dark matter field. However, the density fields will differ in real space, having different residuals when compared to the true dark matter density. The benefit to considering both of these methods is that the linear transfer function is a fully linear approach so it can best demonstrate if there is a need for non-linear methods such as ML.

\subsection{NFW Painting of Individual Halo Profiles}\label{sec:nfwpaint}

Another non-ML approach we consider is painting on NFW profiles using the true halo masses and radii. However, this approach requires known halo parameters and so cannot be applied directly to data without further inference of these parameters. 

The NFW painting is done in Fourier space, which allows us to paint on the profile using the exact halo coordinates. The NFW density profile is given by
\begin{equation}
    \rho(r) = \frac{\rho_s}{\left(r/r_s\right)\left(1+r/r_s\right)^2}
\end{equation}
where $r_s$ is the scale radius which is set by the radius at which $n_{\rm eff}=d\mathrm{ln}\rho/d\mathrm{ln}(r/r_s)=-2$. This can be related to the spherical overdensity definition through the halo concentration $c_\Delta \equiv r_\Delta/r_s$. We use a fit for the concentration parameter from Ref.~\cite{Child_2018} given by
\begin{equation}
    c_{200} = 4.61 \left[ \left(\frac{M_{200}/M_*}{638.65} \right)^{-0.07} \left(1+\frac{M_{200}/M_*}{638.65} \right)^{0.07} -1 \right] + 3.59
\end{equation}
where $M_*=10^{12.5}$. The scale density is given in terms of the spherical overdensity parameters as 
\begin{equation}
    \rho_s = \frac{M_{200}c_{200}^3}{4\pi r_{200}^3} \frac{1}{\mathrm{ln}(1+c_{200})-(c_{200}/(1+c_{200}))}.
\end{equation}
This is fed into the analytic expression for the Fourier transform of the NFW profile \cite{Cooray:2002dia}
\begin{equation}
\begin{split}
    u(k) = & \,4\pi \rho_s r_s^3 \left\{\mathrm{sin}(k r_s)\left[\mathrm{Si}([1+c]kr_s)-\mathrm{Si}(kr_s) \right] - \frac{\mathrm{sin}(ckr_s)}{(1+c)kr_s} \right.\\
    & \left. +\, \mathrm{cos}(kr_s)\left[\mathrm{Ci}([1+c]kr_s) -\mathrm{Ci}(kr_s) \right] \right\}.
    \end{split}
\end{equation}

We define a Fourier-space grid with $k_{x,y,z}=\frac{2\pi}{L}n$ 
where L is the box size, N is the number of grid points per dimension (256), and $n=(0,1,..,N/2,-N/2+1,...,-1)$. The total wavenumber is $k=\sqrt{k_x^2+k_y^2+k_z^2}$. To prevent aliasing, we need to account for the finite cellsize ($L/N$) by convolving with the Fourier transform of the cubic voxel (the 3D sinc function):
\begin{equation}
    \tilde{\rho}(k) = \rho(k) \mathrm{sinc}\left(\frac{k_x L}{N}\right) \mathrm{sinc}\left(\frac{k_y L}{N}\right) \mathrm{sinc}\left(\frac{k_z L}{N}\right).
\end{equation}
We then apply a phase shift to center the halo
\begin{equation}
    \hat{\rho}(k) = \tilde{\rho}(k) e^{-i(k_x x_h + k_y y_h + k_z z_h)}
\end{equation}
where $(x_h,y_h,z_h)$ are the halo's physical coordinates. Finally, we inverse Fourier transform to obtain the real space density
\begin{equation}
    \rho(x,y,z) = \mathcal{F}^{-1}[\hat{\rho}(k)].
\end{equation}
A mask is applied to the field to ensure that all dark matter outside of the halo's $r_{200}$ radius is zeroed out. For $r_{200} < $ cellsize, the mask leaves the dark matter in only that cell, to avoid completely zeroing out subgrid halos.

We compare our results to the true CMD density fields. We consider not just the complete dark matter fields, but also the true field with a mask applied to remove any dark matter outside of the halo's $r_{200}$ radius (ie. the halo field).

\subsection{Halo-Level Machine Learning: GNN--NFW Painting}

This is a ML extension to the NFW painting approach that predicts the halo masses of galaxy groups, allowing application to data. 

\subsubsection{Architecture}
For the GNN part of this method, we use the public GNN code called \texttt{HaloGraphNet}\footnote{\url{https://github.com/PabloVD/HaloGraphNet}}  from Ref.~\cite{Villanueva-Domingo_2022}, which used galaxy properties to infer the mass of halos. The architecture is fully explained in Ref.~\cite{Villanueva-Domingo_2022}, but we summarize it below. The GNN uses message-passing, taking into account the edge ($\mathbf{e_{ij}}$) information through the EdgeNet set-up \cite{wang2019edgenet}:
\begin{multicols}{2}
\begin{enumerate}
\centering
        \item $\mathbf{e^{'}_{ij}} \hookleftarrow \phi_{e}(\mathbf{w_i,w_j-w_i},u)$
        \item $\mathbf{w^{'}_i} \hookleftarrow \mathbf{\mathrm{max}_{j \in \mathcal{N}_i}\mathbf{e{'}_{ij}}}$.
\end{enumerate}
\end{multicols}
\noindent
Here, the node features are $\mathbf{w_i}$ and global features are $u$. The aggregation function is the maximum and $\phi_{e}$ is a MLP with three hidden layers alternated with two ReLU activation functions. The hidden layers respectively have 300, 300, and 100 hidden channels. We use only one message passing layer. To obtain a final graph-level representation, the network employs a combination of global sum, mean, and max pooling. 

\subsubsection{Training}\label{sec:nfwtraining}

We changed the  \texttt{HaloGraphNet} code to have a different dataset format  by creating a complex graph structure. This creates a graph for each halo which contains subgraphs corresponding to the galaxies within that halo. Each node is a galaxy, with the features:
\begin{itemize}
    \item $\vec{x}$ - Galaxy position
    \item $r,g,z$ - Magnitudes of r,g, and z luminosity bands 
    \item $r_{1/2}$ - Stellar half-mass radius
    \item $M_{*}$ - Stellar mass
    \item $\sigma_v$ - Velocity dispersion 
\end{itemize} 
Coordinates are normalized by the sidelength of the box. The stellar half-mass radius and velocity dispersion have a normalization factor applied and the stellar mass is kept in units of $10^{10}\mathrm{M}_\odot/h$. The edges are constructed by linking all nodes within a radius of $2 \ \texttt{Mpc/h}$. The global features are the total number of sub-halos and total stellar mass of the halo. Since we are training on simulations, we already know which galaxies belong to which halos. If this was extended to real data, the galaxies would need to be preprocessed using a group finder algorithm to determine their host halos. Another change from the \texttt{HaloGraphNet} code is that we include halos with no satellite galaxies and apply different selection cuts as outlined in Section~(\ref{sec:dataset}).

The target is the halo mass $M_{200}$ and our dataset is at $z=0$. We use an Adam optimizer with a learning rate 0.001 and a MSE loss
\begin{equation}
    \rm{MSE} = \frac{1}{N} \sum_{i=1}^N (\hat{f_i} - f^{\rm true}_i)^2
\end{equation}
where $\hat{f}$ is the model output and $f^{\rm true}$ is $M_{200}$. We can use the full 27 CV simulations here for training and validation instead of only 24, since we train on the halo catalog not the CMD dark matter field. We use the first 18 for training and the last 9 for validation. The early stopping has a patience of 5 epochs and the minimum improvement in validation loss required to reset the patience counter is 0. The number of training epochs is 100 and the batch size is 4 for training, 1 for validation. 

\subsubsection{GNN-NFW Profile Painting}

After the GNN provides $M_{200}$ predictions, we then paint the dark matter density onto a grid using the NFW profile. This is the same procedure outlined in Section~(\ref{sec:nfwpaint}), but with predictions for the halo mass (and therefore the radius). Note that in this painting procedure, we use the true halo positions. If this was applied to real data the proxy could be the position of the central galaxy, which can be identified for instance by finding the brightest galaxy in the halo.

\subsection{Field-Level Machine Learning: GNN-CNN Approach}

\subsubsection{Architecture}
The GNN-CNN method of inferring the density distribution of dark matter we use here is a modification of the approach introduced in Ref.~\cite{Kvasiuk:2024kwe}. The original implementation consisted of a message-passing GNN block that encodes information from the galaxy feature inputs. Secondly, the GNN output was gridded using a learned grid assignment scheme to create proto-density fields. Lastly, this went into a CNN which acts as the decoder, outputting the density field prediction. In this work, we replaced the learnable grid assignment scheme with the standard CIC procedure for computational simplicity. It resulted in a negligible decrease of the reconstruction quality.

\subsubsection{Training} \label{sec:GNNCNNTraining}
The training dataset is prepared from CAMELS-IllustrisTNG-CV hydrodynamical simulations set as described in Section~(\ref{sec:dataset}). Out of $24$ simulations, we reserve 9 for validation and train our model on the rest. The input data are galaxy features which are the same set listed in Section~(\ref{sec:nfwtraining}). We normalize coordinates by the length of the box and they vary in the unit range. The magnitudes of the luminosity bands are normalized to have zero mean and unit variance. All other features - stellar mass, stellar half-mass radius, and velocity dispersion - are log-normalized, so that $\log_{10}$ of the corresponding quantity has zero mean and variance of one. Scalar features correspond to the graph nodes (e.g. stellar mass), and the translationally invariant graph edges correspond to pairwise differences of vector features (e.g. position). The target is the dark matter and electron overdensities $\rho_m$, $\rho_e$, on a grid of $256^3$ pixels. We scale them to have unit variance and zero mean. The neural network is trained to minimize pixelwise $l_2$ loss with the AdamW optimizer. We set the initial learning rate to $8\times10^{-4}$ and train the model for $100$ epochs with the batch size of $1$, saving the model with the lowest validation loss and reducing the learning rate by a factor of $3$ if the validation loss does not improve for 10 consecutive epochs. The model is trained in mixed precision on a single RTX-3090.

\section{Results: Dark Matter Reconstruction}\label{sec:DMrec}

\subsection{Summary of Results}

\begin{figure}[tbh!]
\centering
  \includegraphics[width=1.05\linewidth]{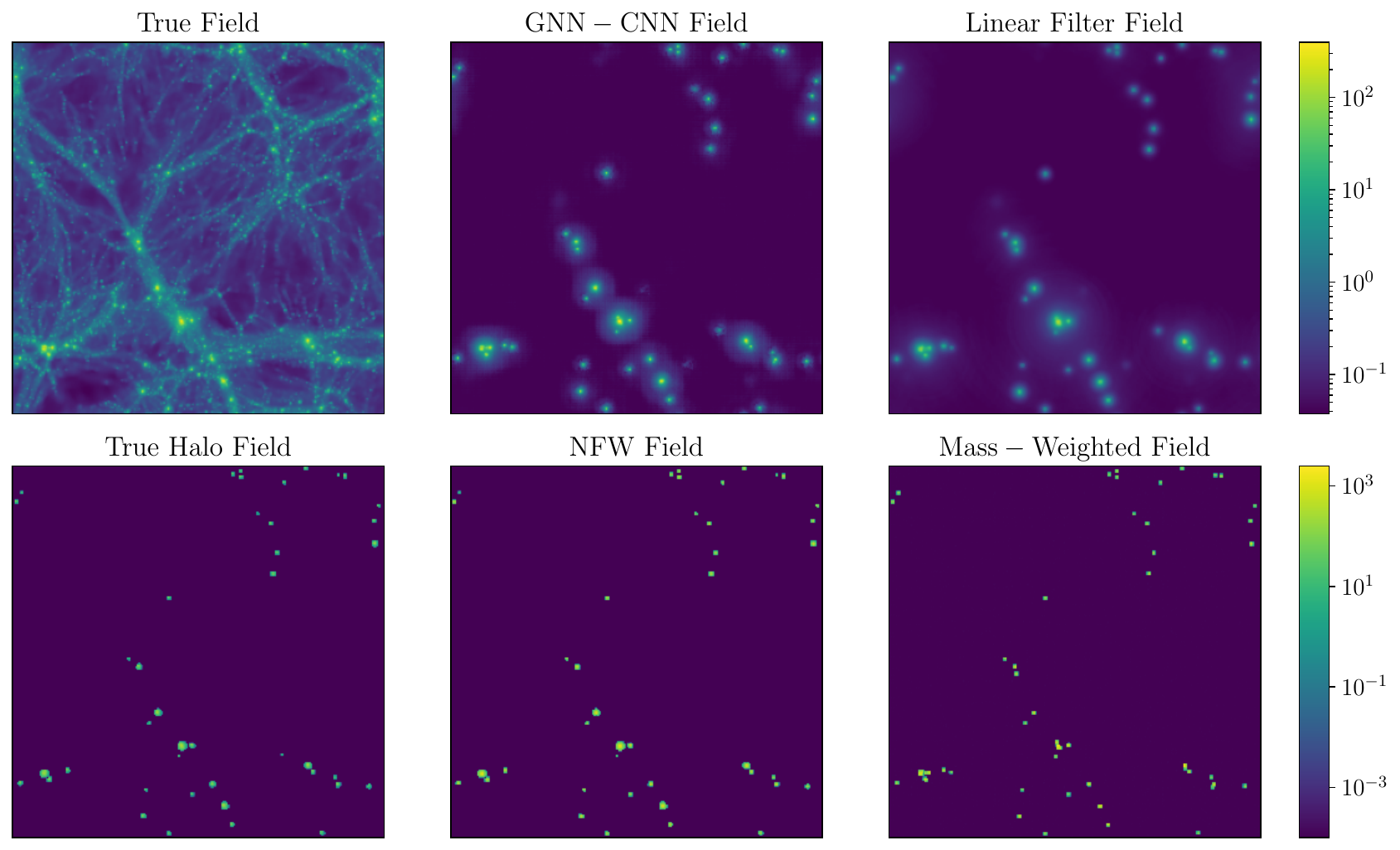}
  \caption{The top row show methods that reproduce the dark matter field and the bottom row shows either the halo or galaxy field. Top row: truth (left), GNN-CNN (middle) and linear filter (right) dark matter density fields. Bottom row: true halo (left), NFW (middle), and mass-weighted density fields. The fields are 1+overdensity, plotted in a $5\times25\times25\ (\texttt{Mpc/h})^3$ volume, averaged over the $x$ axis. The colorbar is logarithmic with values clipped to $10^{-4}$.}
\label{fig:all_model_otp_compare}
\end{figure}

Before going over the results for each of these methods, we provide an overview of their performance. In Fig.~(\ref{fig:all_model_otp_compare}), a slice of the density field is shown for each method. The left column is the true dark matter field (top) and the true halo field (bottom). The latter is defined as only the dark matter within the radius of the halo that encloses 200 times the critical density ($r_{200}$). Methods in the top row reproduce the dark matter field, but in the bottom row they either construct the halo field (NFW field) or the galaxy field as a biased tracer of the halos (mass-weighted field). A limitation of all these methods is their reliance on discrete tracers, meaning many areas of the cosmic web cannot be reconstructed. However, these regions are less crucial because the density is much lower, the log-scale of the plots artificially amplifies their significance.

\begin{figure}[h!]
\centering
  \includegraphics[width=\linewidth]{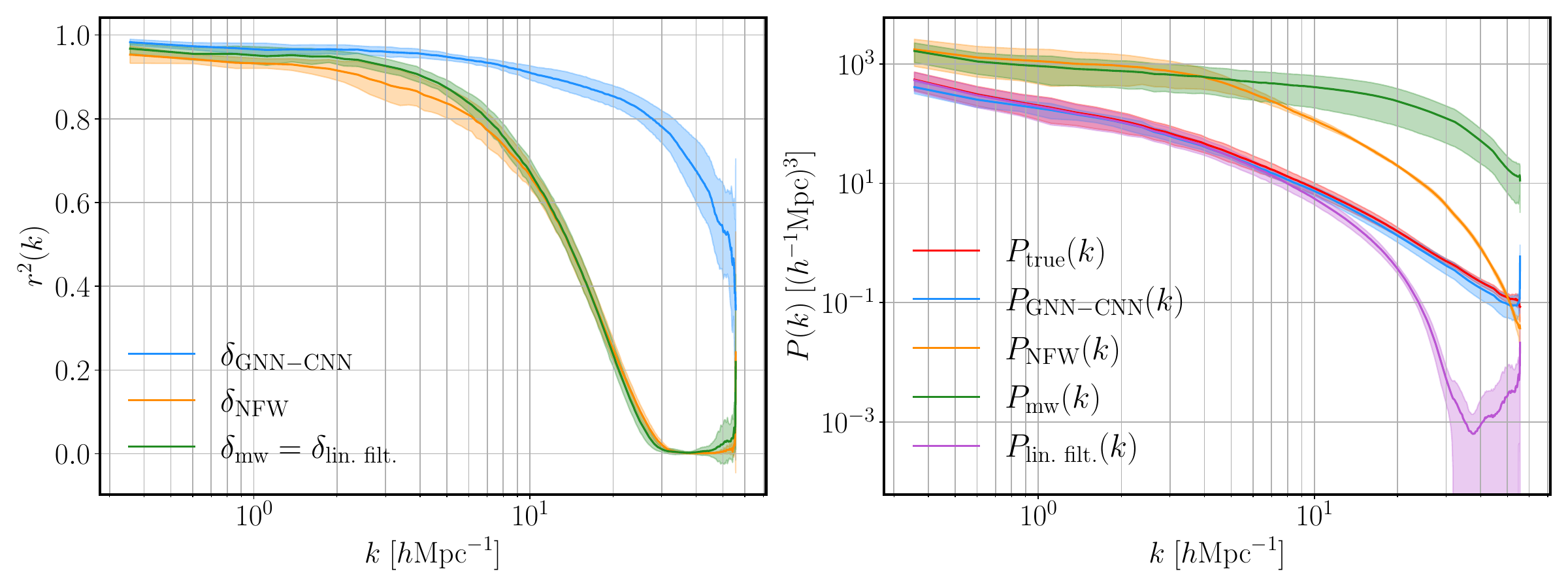}
  \caption{Left: The cross-correlation coefficients between the true dark matter density field and reconstructed fields. The reconstructed fields are the GNN-CNN (blue), NFW (orange), and galaxy mass-weighted/linear filter  (green) fields. Right: The power spectrum of the true dark matter field (red) compared to GNN-CNN (blue), NFW (orange), galaxy mass-weighted (green) and linear filter (purple) fields. The NFW (mass-weighted) field reproduces the halo (galaxy) $P(k)$ so it is related to $P_\mathrm{true}(k)$ by the halo (galaxy) bias. }
  \label{fig:rk_pk_dm_compareall}
\end{figure}

Quantitatively, when compared to the true dark matter field, as shown by the cross-correlation coefficients in Fig.~(\ref{fig:rk_pk_dm_compareall}), the GNN-CNN consistently achieves the best performance, demonstrating improved correlation across all scales. Both the linear transfer function and the mass-weighted field (our baseline method, which grids the stellar density at observed galaxy locations) have the same cross-correlation coefficient with the true dark matter field. While the NFW field effectively paints on density profiles at halo locations and performs well at reproducing the halo field, it does not perform as well as the other methods when reconstructing the full dark matter field. The mass-weighted and linear transfer function fields account for dark matter dispersed outside of halos by also using sub-halo locations, so they do better than the NFW approach at reconstructing the full dark matter distribution. 

We also reconstruct the gas field, which is shown in Section~(\ref{sec:gas}). The GNN-CNN again had the best reconstruction. We did not consider more cases in this work, but any field given by the simulations could be learned by the GNN-CNN architecture. 

\subsection{Baseline: Linear Transfer Function}

The linear transfer function method results in the same cross-correlation coefficient as the galaxy mass-weighted field, as shown in Fig.~(\ref{fig:rk_pk_dm_compareall}). The power spectrum from the  linear transfer function matches the true dark matter field spectrum until large $k$. A log-scale slice of the density field is illustrated in Fig.~(\ref{fig:all_model_otp_compare}). The linear transfer function paints on the density to larger radii while the mass-weighted field are higher density in smaller regions. The linear transfer function field does not capture the small-scale structure as well as the non-linear technique of the GNN-CNN method as we will see below. 

\subsection{NFW Painting of Individual Halo Profiles}\label{sec:NFWresults}

\begin{figure}[h!]
\centering
  \includegraphics[width=\linewidth]{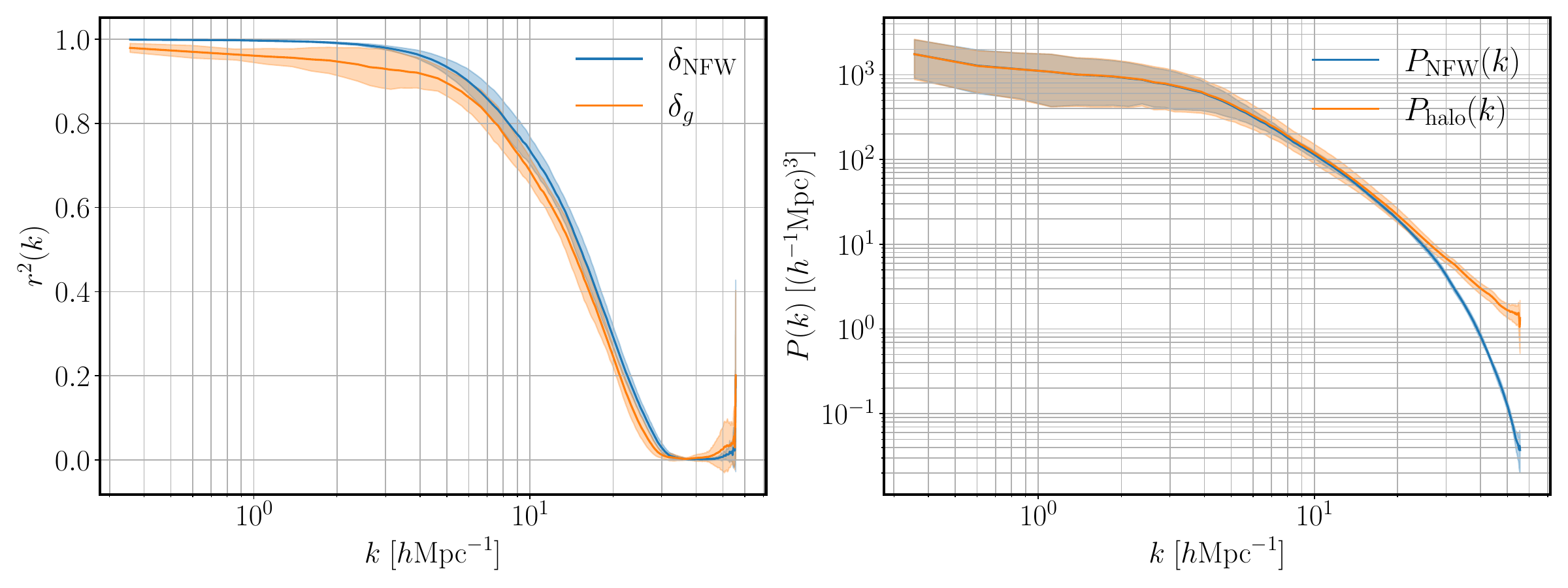}
  \caption{Left: The cross-correlation coefficients between the true halo field and the NFW painted dark matter density, $\delta_{\rm NFW}$ (blue), and the galaxy mass field $\delta_{g}$ (orange). Right: The $P(k)$ of the true halo field (orange) and the $P(k)$ of the NFW density field (blue). $P_{\rm halo}(k)$ is rescaled to match $P_{\rm GNN-NFW}(k)$.}
  \label{fig:rk_dm_nfw_masknotall}
\end{figure}

The NFW method paints on halo profiles, so it reproduces the halo power spectrum (over a restricted range of halo masses) rather than the full dark matter power spectrum. To make an accurate comparison of the NFW field to the true dark matter field, we mask the latter to only include regions inside of halos. Fig.~(\ref{fig:rk_dm_nfw_masknotall}) compares both the NFW painted field and the mass-weighted fields to the masked true halo field. The left panel shows the correlation coefficients. It is clear that the NFW field outperforms the mass-weighted field on all scales, even achieving perfect correlation at large-scales ($r(k) = 1$ at low $k$). In the right panel of the figure, we compare the power spectra. By masking the true dark matter field, we now have a true halo field $P(k)$, but it has a different amplitude because masking removes matter. To account for this, we have rescaled the $P_\mathrm{halo}(k)$ to match $P_{\rm NFW}(k)$ at low $k$. After rescaling, the power spectra match well down to small-scales. 

Next, we tested how NFW painting performs compared to the full dark matter field, not just the halo component. As shown in Fig.~(\ref{fig:rk_pk_dm_compareall}), for small-intermediate $k$ the NFW field is less correlated with the truth than performing CIC on the galaxy stellar masses. The right panel depicts the power spectra. Since the NFW field is discrete, it does not have the same shape as the dark matter power spectrum, and they are offset on large-scales by the halo bias.

\begin{figure}[tbh!]
\centering
  \includegraphics[width=1.\linewidth]{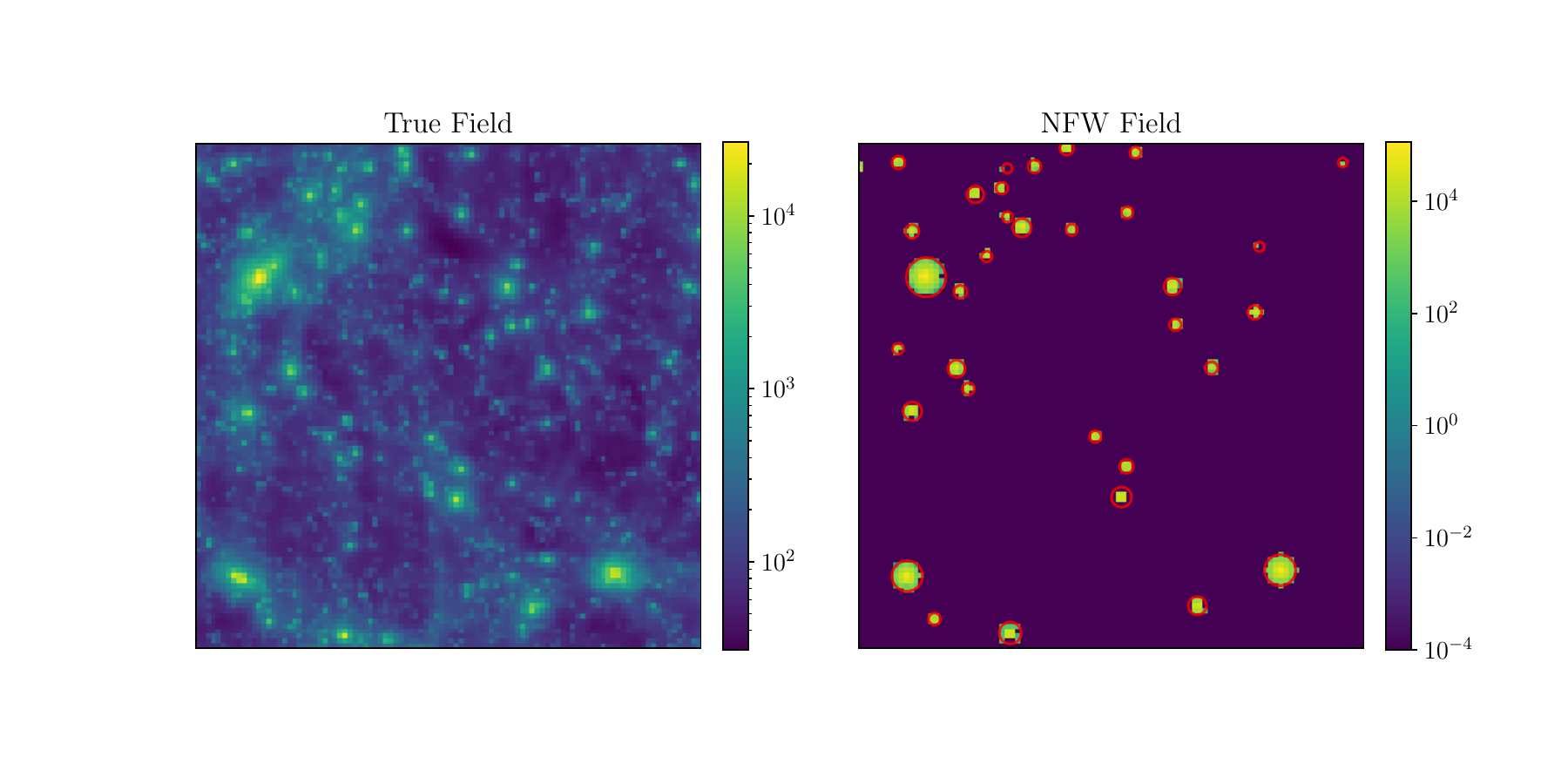}
  \caption{True (left) and predicted (right) NFW dark matter density, with red circles of $r_{200}$ radius. The images are summed over the x-axis and zoomed in to a $10\times10\ (\texttt{Mpc/h})^2$ region. These are plotted as 1 + overdensity and clipped to $10^{-4}$ for the logarithmic colorscale.}
\label{fig:NFW_model_circle_compare}
\end{figure}

A realspace comparison can provide a better picture of how the fields differ. In Fig.~(\ref{fig:all_model_otp_compare}) is a zoomed-in log-scale slice of the density fields. In the bottom row, we see that the true halo field looks similar to the NFW painted field. The NFW painted field has placed spherical density profiles over the halos that extend to $r_{200}$. These profiles extend farther out than the mass-weighted field. Another difference is that the mass-weighted field is painting on the location of each of the galaxies in our catalog, instead of on the location of halos. The NFW and mass-weighted fields conserve the mass of the simulation, while only painting on discrete objects. This results in halo densities that are higher than those seen in the true field. The top left image of the true full dark matter field shows filamentary structures that cannot be reproduced by the NFW or mass-weighted methods. To examine the difference in more detail, Fig.~(\ref{fig:NFW_model_circle_compare}) is a zoomed in 2D projection of the true dark matter and NFW fields. The NFW field has red circles that show the $r_{200}$ at which the density is cutoff.

\begin{figure}[htbp]
    \centering
    \includegraphics[width=\linewidth]{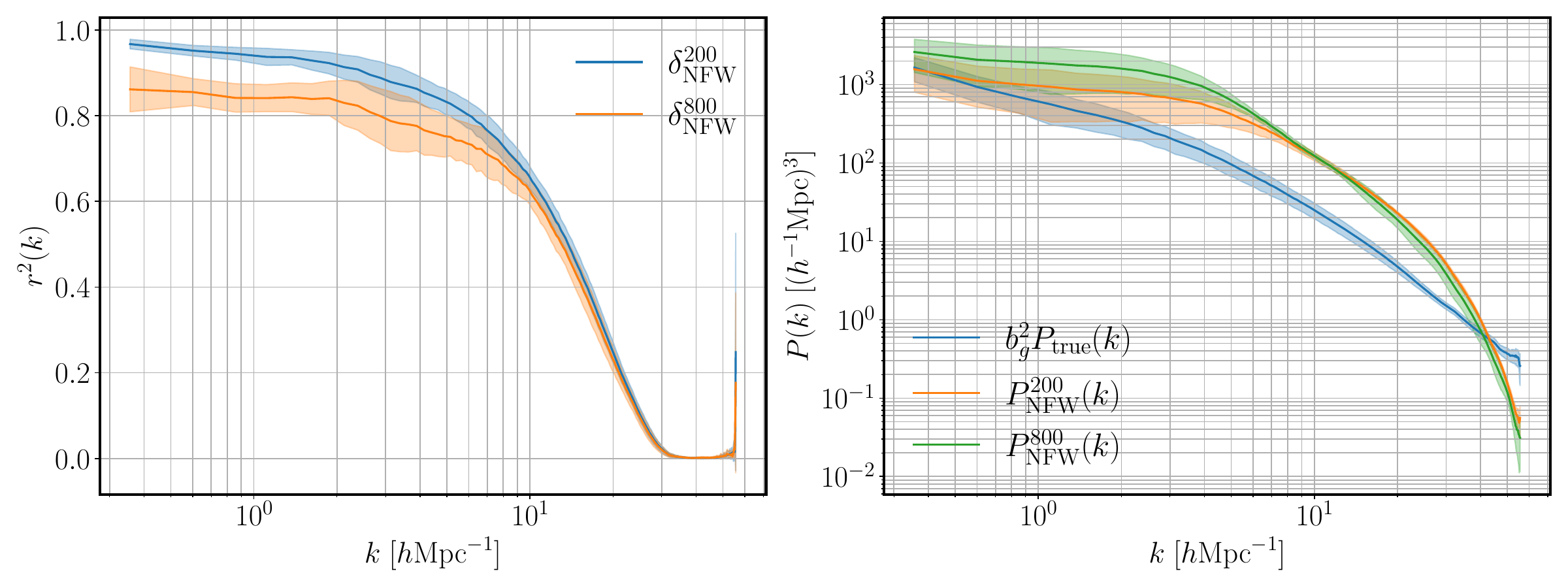}
    \includegraphics[width=\linewidth]{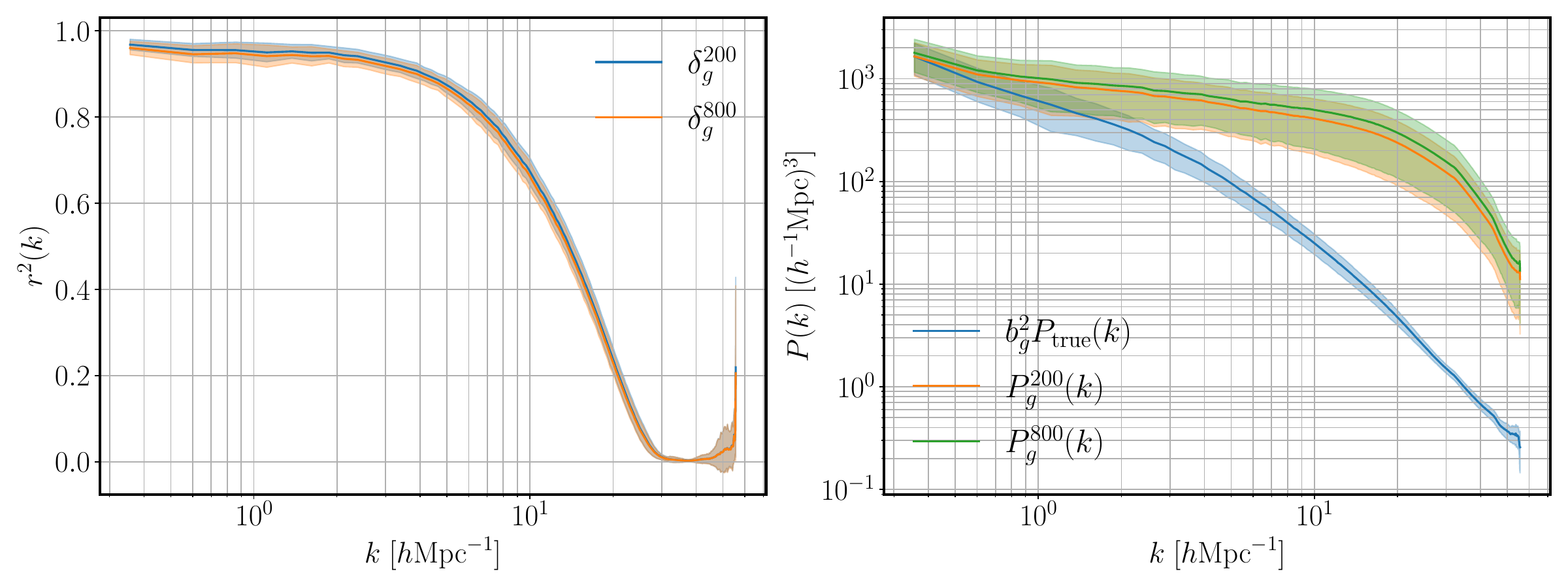}

    \caption{Left: The cross-correlation coefficient of the true dark matter density field with the NFW (top) or mass-weighted field (bottom). In blue are the results using the $n_{\rm part}>200$ selection cut and orange
    is $n_{\rm part}>800$. Right: The $P(k)$ of the true dark matter field rescaled by $b_g\mid_{n_{\rm part}>200}$ (blue). In the top image is the NFW painted density field $P(k)$ for $n_{\rm part}>200$ in orange and $n_{\rm part}>800$ in green. The bottom image has the same cases but for the mass-weighted field.}
    \label{fig:combined_rk_pk_dm} 
\end{figure}

A final check is how the NFW method changes with the galaxy number density, compared to the galaxy mass-weighted approach. When creating our galaxy catalog, one of the selection criteria we applied was that the number of stellar and dark matter particles ($n_{\rm part}$) must be greater than 200, which resulted in an average of 283 galaxies in a simulation. We increased this threshold to $n_{\rm part}>800$ which changed the average number of galaxies to 136. This allows us to compare how these methods would perform on a galaxy survey with a lower number density of objects. As seen in the top panel of Fig.~(\ref{fig:combined_rk_pk_dm}) for the NFW field and in the bottom panel for the galaxy mass-weighted field, the correlation coefficient decreases with the number density of galaxies. This is more pronounced for the NFW field. To enable a rough comparison of the power spectra, we scaled the true dark-matter power spectrum by the galaxy bias to approximate the halo clustering signal (i.e., the 2-halo term). However, this only provides a comparison at the lowest $k$. We used the galaxy bias from the $n_{\rm part}>200$ galaxy catalog. The mass-weighted field $P(k)$ does not match the true dark matter field because it is the galaxy power spectrum. 

These results demonstrate that NFW painting can outperform the stellar mass-weighted approach if our goal is to reconstruct the halo field. If the goal is to reconstruct the full dark matter field, then mass-weighting is better. It is important to note that in practice we do not know the true halo parameter values, so this idealized scenario cannot be applied to real data. Furthermore, we used halos found from particles for the training dataset and use the true halo positions from the simulations. If this were done more realistically, we would use halos found from grouping galaxies and the central galaxy's position as a proxy for the halo's position. These changes would further lower the performance, suggesting that the NFW painting approach is suboptimal to a galaxy mass-weighted field. In light of this result, we do not extend a halo-painting approach to gas reconstruction. 

\subsection{Halo-Level Machine Learning: GNN-NFW}

\begin{figure}[h!]
\centering
\includegraphics[width=0.6\linewidth]{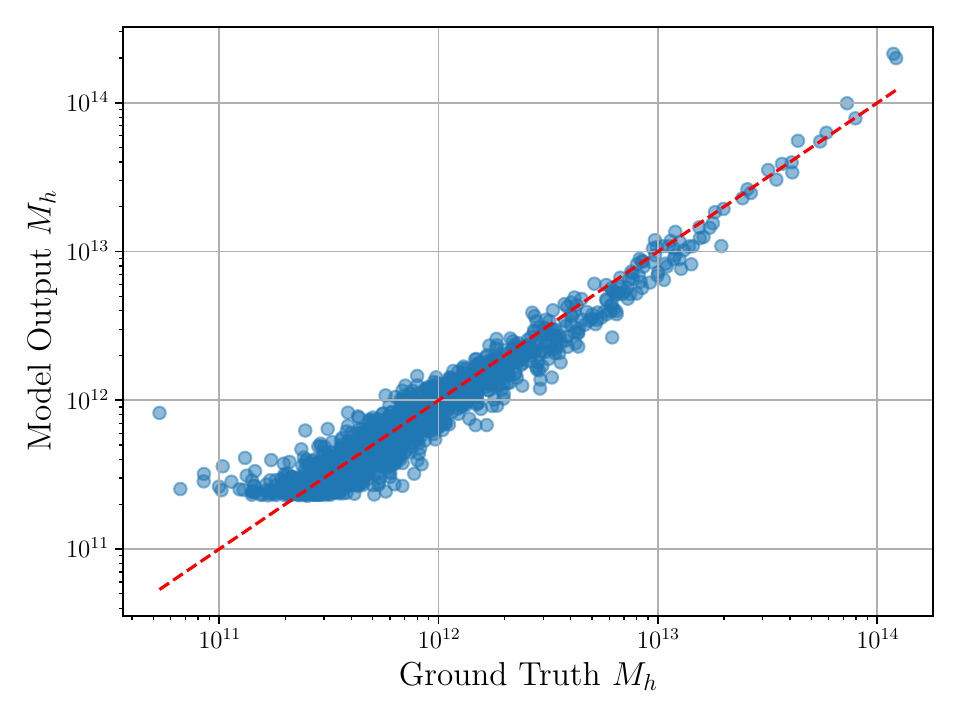}
  \caption{The GNN model output for halo mass compared to the true values. The red dashed line represents perfect correlation.}
  \label{fig:GNN_preds}
\end{figure}

We now explore how the performance of halo painting is affected when a GNN is used to predict the halo properties used in the NFW profiles from galaxy features. Fig.~(\ref{fig:GNN_preds}) is a comparison of the outputs of the GNN model with the ground truth for $M_{200}$. The red dashed line shows where the model outputs correspond exactly to the true values. From this we can see that the GNN performs well at intermediate $M_{200}$, but deviates more at the lowest and highest masses. This makes sense as there are fewer halos in these regions to train on, especially at the high-mass end. The model achieves a scatter of $\sim 0.02$ dex in $\mathrm{log}_{10}(M / (10^{10}M_\odot/h))$.

\begin{figure}[htbp]
    \centering
    \includegraphics[width=\linewidth]{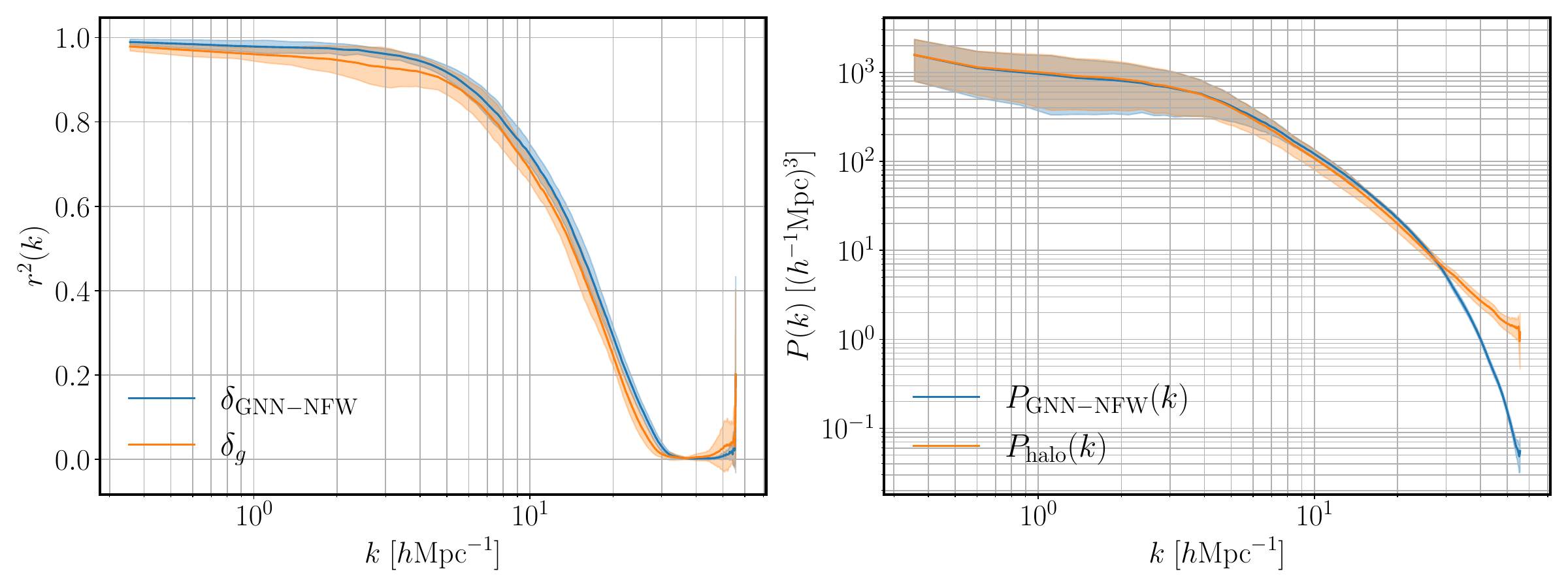}
    \includegraphics[width=\linewidth]{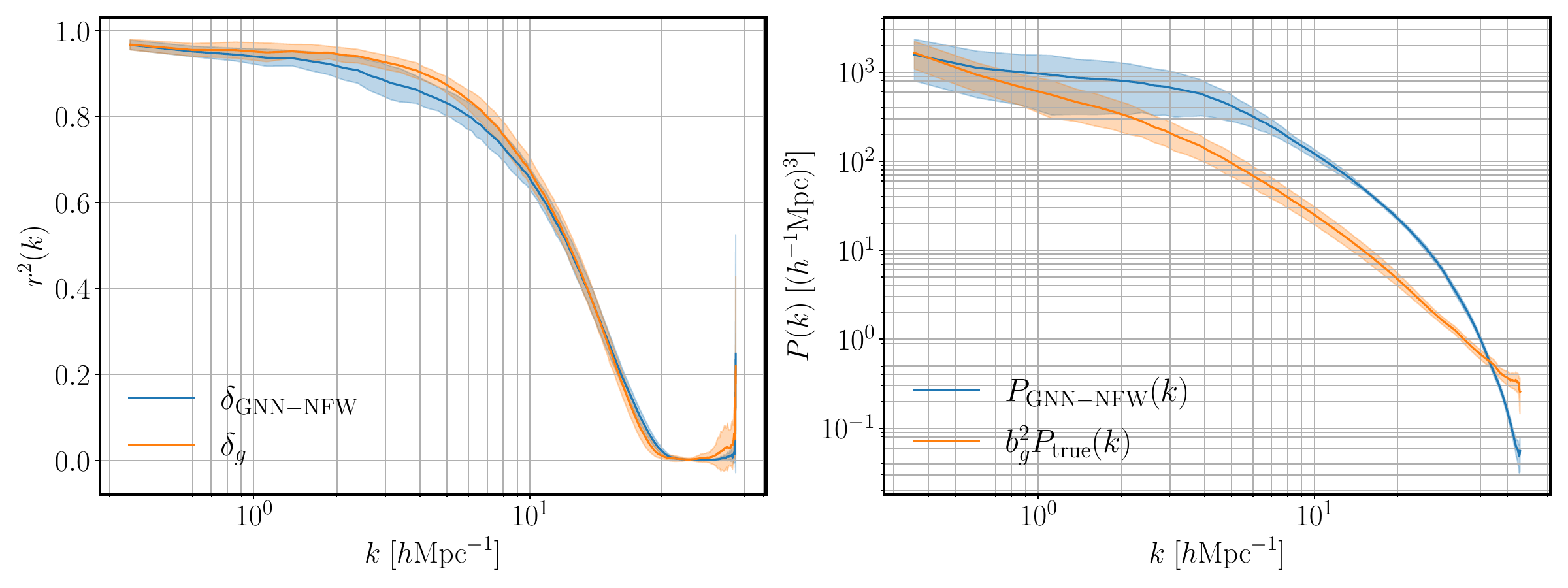}

    \caption{Left Top: The cross-correlation coefficient of the GNN-NFW painted dark matter density, $\delta_{\rm NFW}$ (blue), and galaxy mass field $\delta_{g}$ (orange) with a true halo field. Right Top: The $P(k)$ of the true halo field (orange) and the $P(k)$ of the GNN-NFW reconstructed density field (blue). $P_{\rm halo}(k)$ is rescaled to match $P_{\rm GNN-NFW}(k)$. Bottom: Same images as the top, but for the true dark matter field. The true dark matter density's $P(k)$ is rescaled by galaxy bias squared.}
    \label{fig:combined_rk_dm_nfw_pred} 
\end{figure}

The top left panel of Fig.~(\ref{fig:combined_rk_dm_nfw_pred}) shows the correlation coefficients of the true halo field with both the GNN-NFW painted field and the mass-weighted field. Across all scales, 
the true field correlates better with the  $\delta_{\rm GNN-NFW}$ field than the $\delta_g$ field. However, the GNN-NFW field has a lower cross-correlation coefficient than the NFW painted field shown in Section~(\ref{sec:NFWresults}). This is to be expected since the GNN-NFW method accounts for our uncertainty in the halo masses. In the right panel, we see that the power spectrum aligns well with the true halo field until small-scales, but the NFW field had slightly better agreement at intermediate scales.

Fig.~(\ref{fig:combined_rk_dm_nfw_pred}) compares GNN-NFW to the full dark matter field in the bottom panel. Overall, GNN-NFW has a similar performance as the NFW painted case (Fig.~(\ref{fig:rk_pk_dm_compareall})). On large-scales, GNN-NFW shows slightly higher correlation, but this is likely coincidental, as both methods aim to reconstruct the halo field, not the full dark matter field. The GNN-NFW power spectrum also closely resembles the NFW result. Not depicted here is the GNN-NFW field in realspace since visually it has the same behaviour as the NFW field. 

Overall, the results mirror what we saw for NFW painting, but this approach has some realism added to account for our uncertainty about the halo mass values. The conclusion is the same, that NFW painting based methods can improve upon stellar mass-weighting for the halo field, but not for the full dark matter field. 

\subsection{Field-Level Machine Learning: GNN-CNN}

\begin{figure}[tbh!]
\centering
  \includegraphics[width=1.\linewidth]{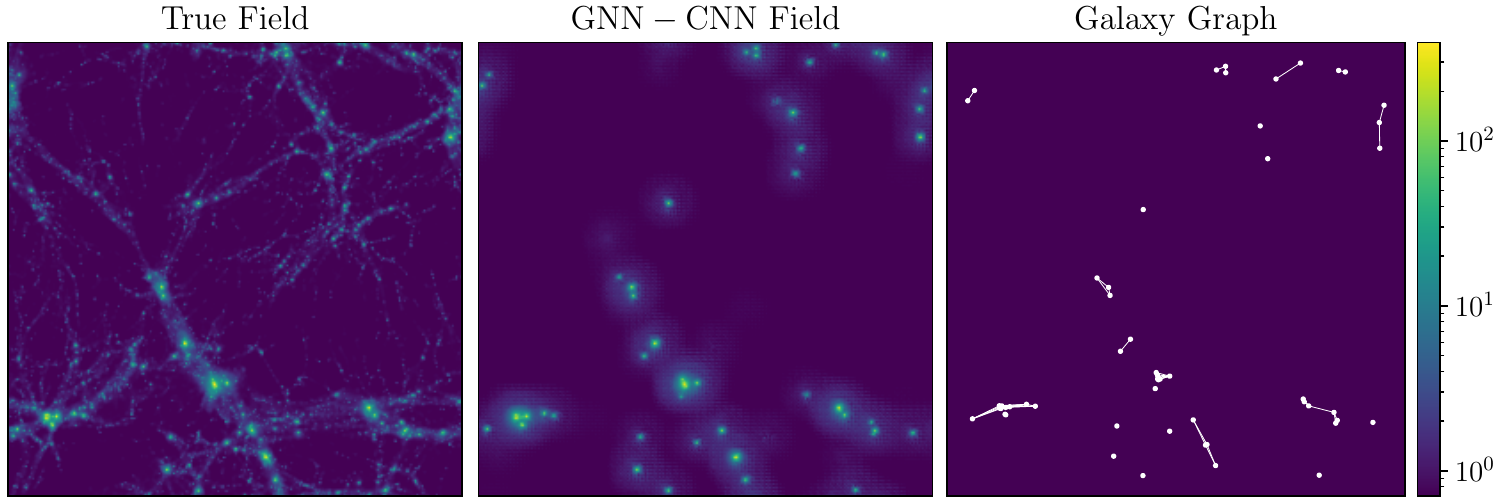}
  \caption{True (left) and predicted with GNN-CNN setup (middle) dark matter density in a $5\times25\times25\ (\texttt{Mpc/h})^3$ volume, averaged over the $x$ axis. The rightmost column shows the visualization of the input galaxy cloud, in the same region, as a graph.}
\label{fig:gnncnn_dm_model_output}
\end{figure}

\begin{figure}[h!]
\centering
\includegraphics[width=1.\linewidth]{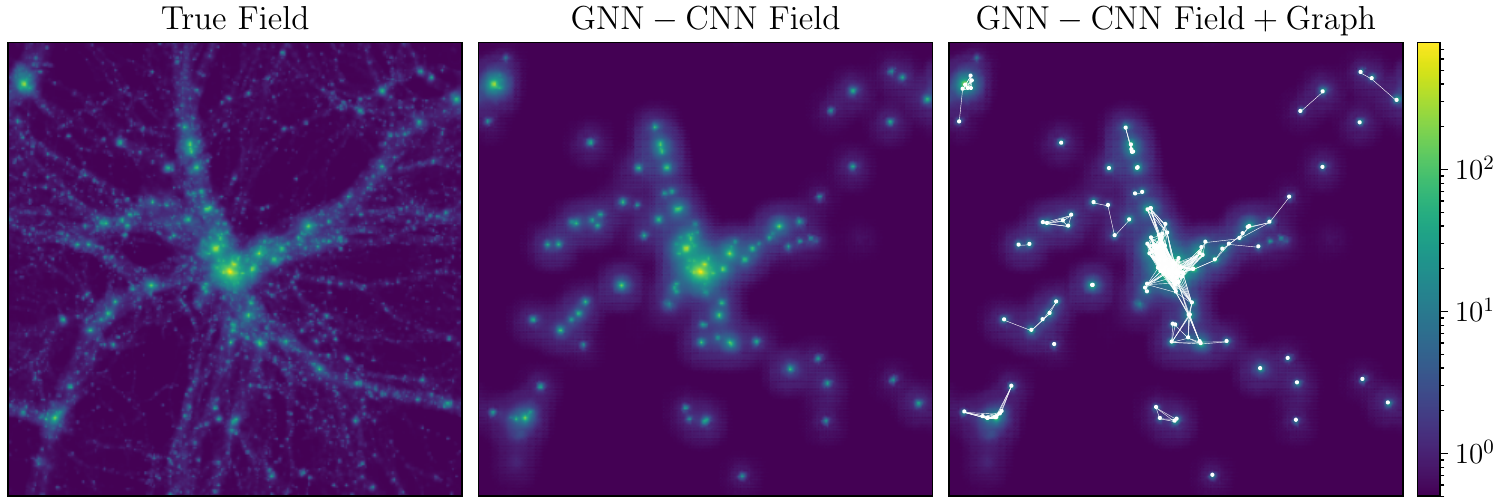}
  \caption{Example of the reconstructed dark matter density $\rho/\bar{\rho}$ (middle), in the $25\times25\times10\ (\texttt{Mpc/h})^3$ volume, projected along the $z$ axis. The right plot shows the overlaid input galaxy graph, and the plot on the left is the target density.}
  \label{fig:gnncnn_model_output_compare1}
\end{figure}

\begin{figure}[h!]
\centering
  \includegraphics[width=\linewidth]{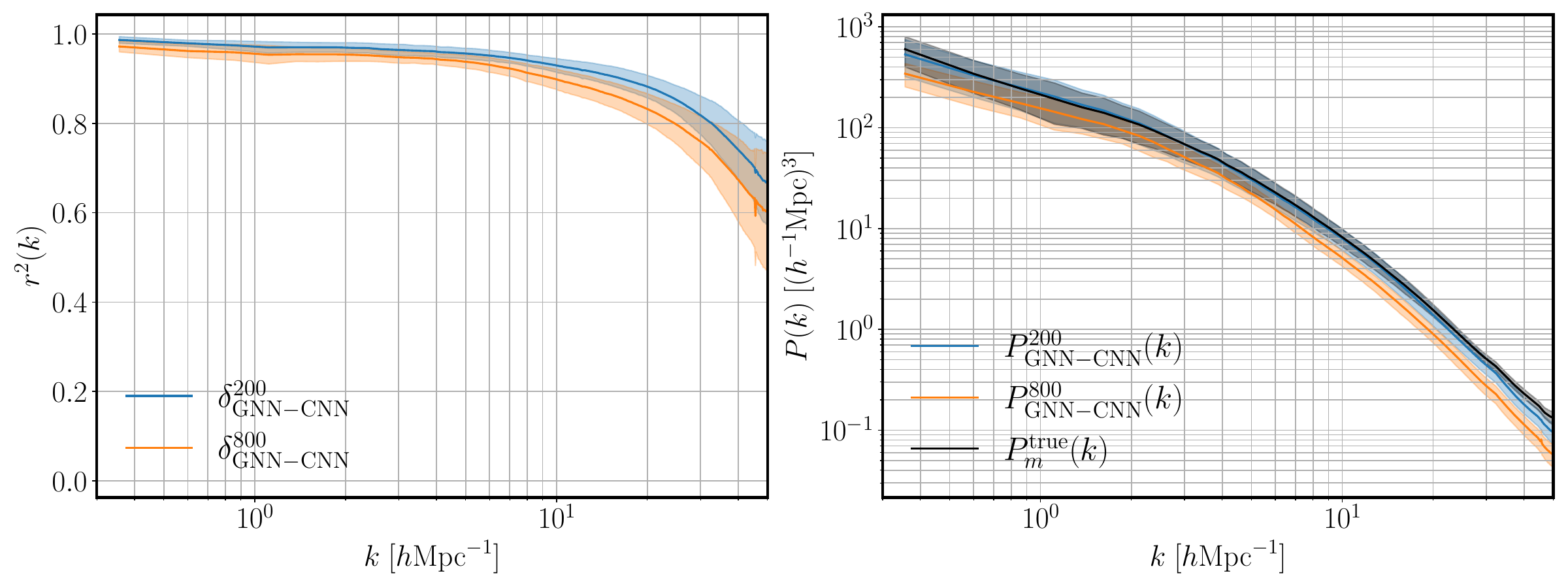}
  \caption{Cross correlation coefficients and power spectra of reconstructed dark matter density fields with GNN-CNNs trained on different number densities. Cut off of 200 particles per galaxy corresponds to 283 galaxies per volume on average (blue), while 800 particles results in 136 galaxies (orange).}
  \label{fig:rk_pk_dm_npart}
\end{figure}

We show the cross correlation coefficient of the dark matter overdensity, reconstructed with the  GNN-CNN setup (blue), and the mass-weighted galaxy field (green) with the true dark matter overdensity in Fig.~(\ref{fig:rk_pk_dm_compareall}) (left). The right plot shows the true and recovered power spectra. As can be seen, the method provides a high-quality and unbiased reconstruction up to the highest wave numbers, for example $r^2(k)\big|_{k=3\ \texttt{Mpc/h}}\approx 0.8$. It is also worthwhile to look at the reconstructed fields. Fig.~(\ref{fig:gnncnn_dm_model_output}) shows the true (left) and reconstructed (middle) $\delta_m$ along with the input galaxy catalog, visualized as a graph (right). We take a slice of the 3D field with a width of $5\ \texttt{Mpc/h}$ and project it along the slice axis. We see that in the regions where we observe a lot of galaxies, the model provides a good reconstruction of the underlying dark matter field. The model struggles with more complex structures, like sheets and filaments. This is expected since there are not so many objects to learn from in such regions. We also show the reconstructed field in a more dense region in Fig.~(\ref{fig:gnncnn_model_output_compare1}). To check the robustness of our method to the number density of observed galaxies, we repeat the training, imposing the particle number cut. In our default configuration, we required that each galaxy consists of at least 200 stellar particles. We now raise this requirement to 800 particles. This results in a reduction of the average number of galaxies per simulation from 283 to 136. The resulting power spectra and cross-correlation coefficients are shown in Fig.~(\ref{fig:rk_pk_dm_npart}). We can see that the cross-correlation coefficient decreases slightly at higher $k$ values. We also notice that the reconstructed field with the higher particle number cut contains less power at all scales. It is natural to expect, since our method paints a dark matter field around each individual object, and fewer objects will result in lower variance on average.   

\section{Results: Gas Reconstruction}\label{sec:gas}

We now move on to describe the reconstruction of the gas field. For NFW/GNN-NFW painting, the results would be significantly worse than for dark matter, since gas is more diffuse, so less of it is located in the halo component of the gas field. Similarly for the mass-weighted field, galaxies will not capture the extended gas profiles. Therefore, we present only the gas results for the GNN-CNN and linear transfer function methods. 

\begin{figure}[h!]
\centering
  \includegraphics[width=\linewidth]{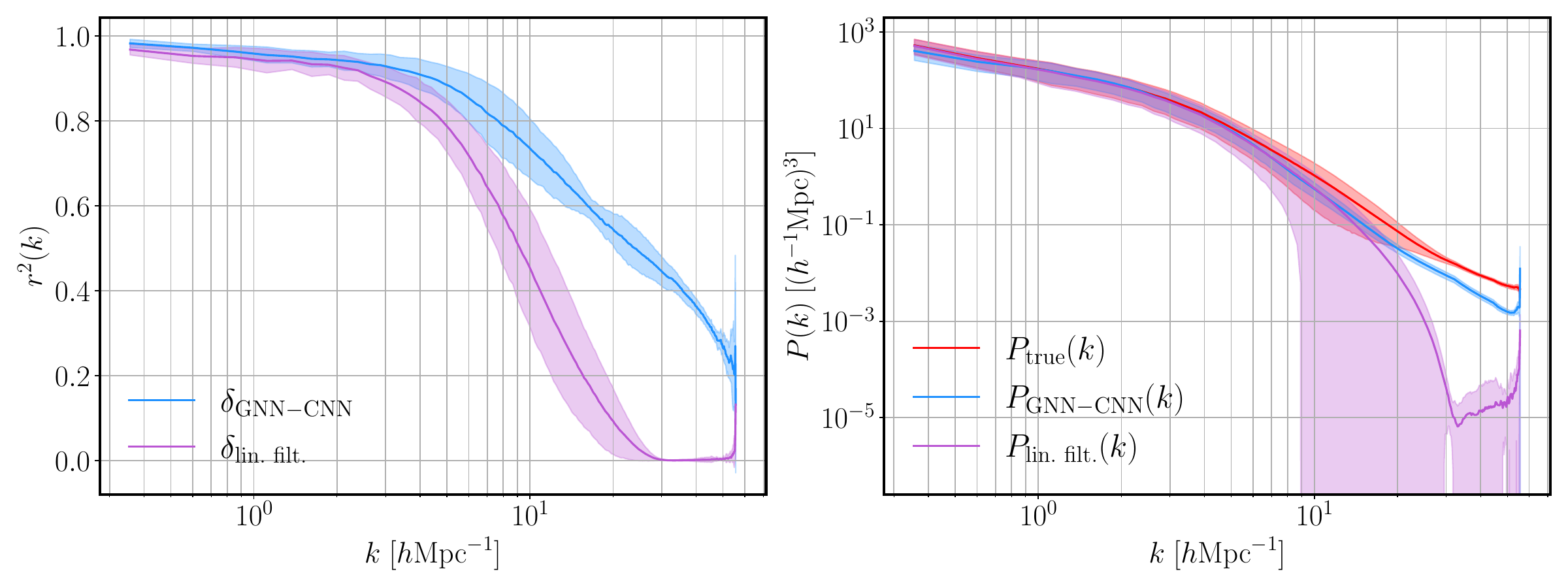}
  \caption{Left: The cross-correlation coefficients between the true gas density field and GNN-CNN (blue) and  linear transfer function (purple) reconstructed fields. Right: The true gas field (red) compared to GNN-CNN (blue) and  linear transfer function (purple) fields. }
  \label{fig:rk_pk_gas_compareall}
\end{figure}

\begin{figure}[tbh!]
\centering
  \includegraphics[width=1.\linewidth]{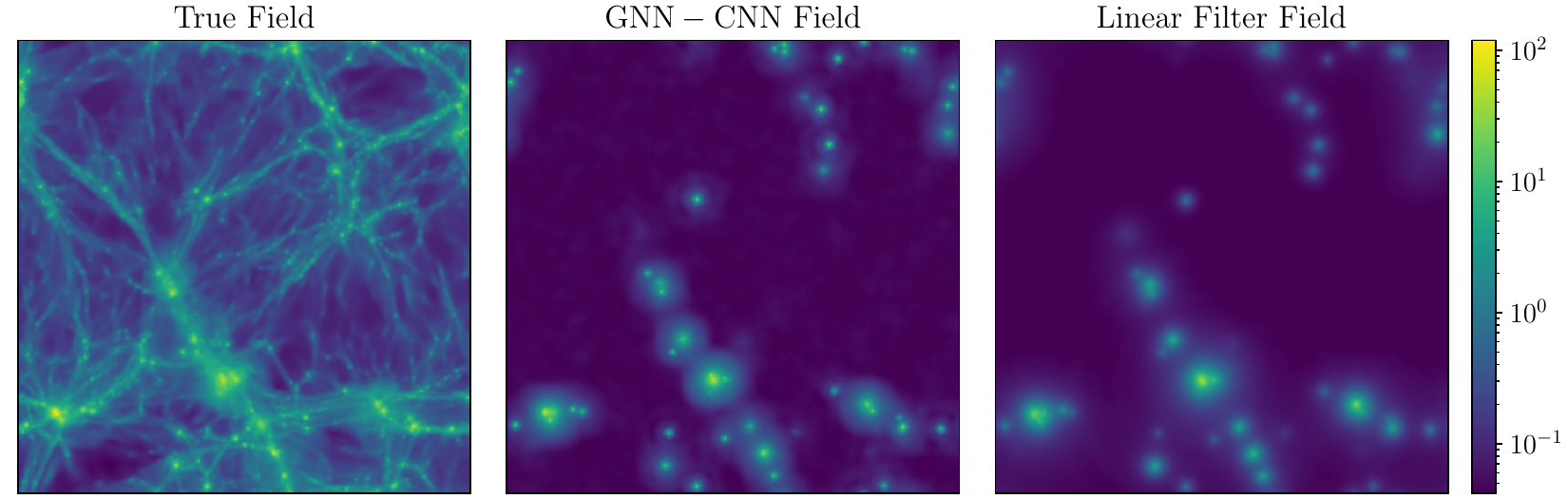}
  \caption{The 1$+$overdensity gas fields in a $5\times25\times25\ (\texttt{Mpc/h})^3$ volume, averaged over the $x$ axis. On the left is the truth, middle is the GNN-CNN and right is  linear transfer function gas field. The colorbar is logarithmic with values clipped to $10^{-4}$.}
\label{fig:all_gas_model_otp_compare}
\end{figure}

The cross-correlation coefficients in Fig.~(\ref{fig:rk_pk_gas_compareall}) demonstrate that the GNN-CNN outperforms the linear transfer function on all scales. Small-scales are significantly improved by using this ML approach, which can also be seen at the level of the power spectra in the right-hand panel of Fig.~(\ref{fig:rk_pk_gas_compareall}). At the field level, the two methods are similar, but the linear transfer function field is a bit more diffuse - as shown in Fig.~(\ref{fig:all_gas_model_otp_compare}). The true field image also illustrates how the majority of gas is outside halos, as opposed to dark matter (Fig.~(\ref{fig:all_model_otp_compare})), which highlights the importance of field-based reconstruction. 

This result is important for kinetic Sunyaev-Zel'dovich (kSZ) velocity reconstruction \cite{Ho:2009iw,Zhang_2010,Shao_2011,Munshi:2015anr,Deutsch:2017_2Dksz_estim1,Smith:2018bpn,Cayuso:2021ljq,McCarthy:2024_ACT_DESI_ksz,Bloch:2024PlanckxUnwise,Lague:2024czc_7.2sigma,Krywonos:2024mpb,Lai:2025qdw,Hotinli:2025tul}. The kSZ effect is a cosmic microwave background (CMB) secondary anisotropy contribution from Thomson scattering of CMB photons by free electrons in bulk-motion ~\cite{Sunyaev1980}. It is proportional to the optical depth and velocity. To extract a velocity from the kSZ effect, a model for the optical depth is needed. Currently, the linear transfer function method is used to model the electrons from observed galaxies. Therefore, the small-scale improvement shown by the GNN-CNN could improve kSZ velocity reconstruction. This idea was introduced in \cite{Kvasiuk:2023nje}, where it was shown that the velocity quadratic estimator noise is inversely proportional to the integrated cross-correlation coefficient of the template with the true electron field, so the improvement in $r^2(k)$ directly translates into the improvement in the velocity reconstruction. For this application, modeling uncertainties create a scale-independent bias (on large-scales) (eg. Refs.~\cite{Battaglia:2016xbi,Smith:2018bpn,Madhavacheril:2019buy,Giri:2020pkk,Cayuso:2021ljq}) so this can absorb any bias from our training data differing from real data. 

\section{Robustness Against Unknown Feedback Parameters}\label{sec:robust}

We now discuss the robustness of the machine-learned dark matter field to changes in the astrophysical feedback parameters. In all previous discussions, the model was trained and tested on simulations from the same set with the same fiducial values of cosmological and astrophysical parameters. We now relax this assumption and discuss generalization outside of the training data. In our previous work, Ref.~\cite{Kvasiuk:2024kwe}, we explored the sensitivity of the GNN-CNN to training on simulations with varied cosmological and astrophysical parameters. We found that the cross-correlation coefficients were unchanged whether it was trained conditionally or unconditionally on the parameter values. In this paper however, we focus only on varying astrophysical parameters because we have precise measurements of cosmological parameters from data, whereas astrophysics is more uncertain - recent kSZ data shows that feedback is much stronger than our simulations predict e.g. \cite{10.1093/mnras/stae2100}. Therefore, in the following we test how the reconstruction results are biased if the fiducial astrophysics differs from reality.

\paragraph{Performance on simulations with different astrophysical parameters.} It is interesting to see how the model performs on a simulation that has significantly different fiducial astrophysical parameter values. For this purpose, we use the $\texttt{EX}$ subset of CAMELS-IllustrisTNG. This subset has 4 simulations, one with fiducial values and 3 with the values of the astrophysical parameters at their extremes (but fixed cosmological parameters). We consider three cases: 
\begin{itemize}
    \item Fiducial: 282 galaxies
    \item Efficient Active Galactic Nuclei (AGN) feedback: 284 galaxies
    \item No feedback: 1018 galaxies.
\end{itemize}

\begin{figure}[h!]
\centering
  \includegraphics[width=0.7\linewidth]{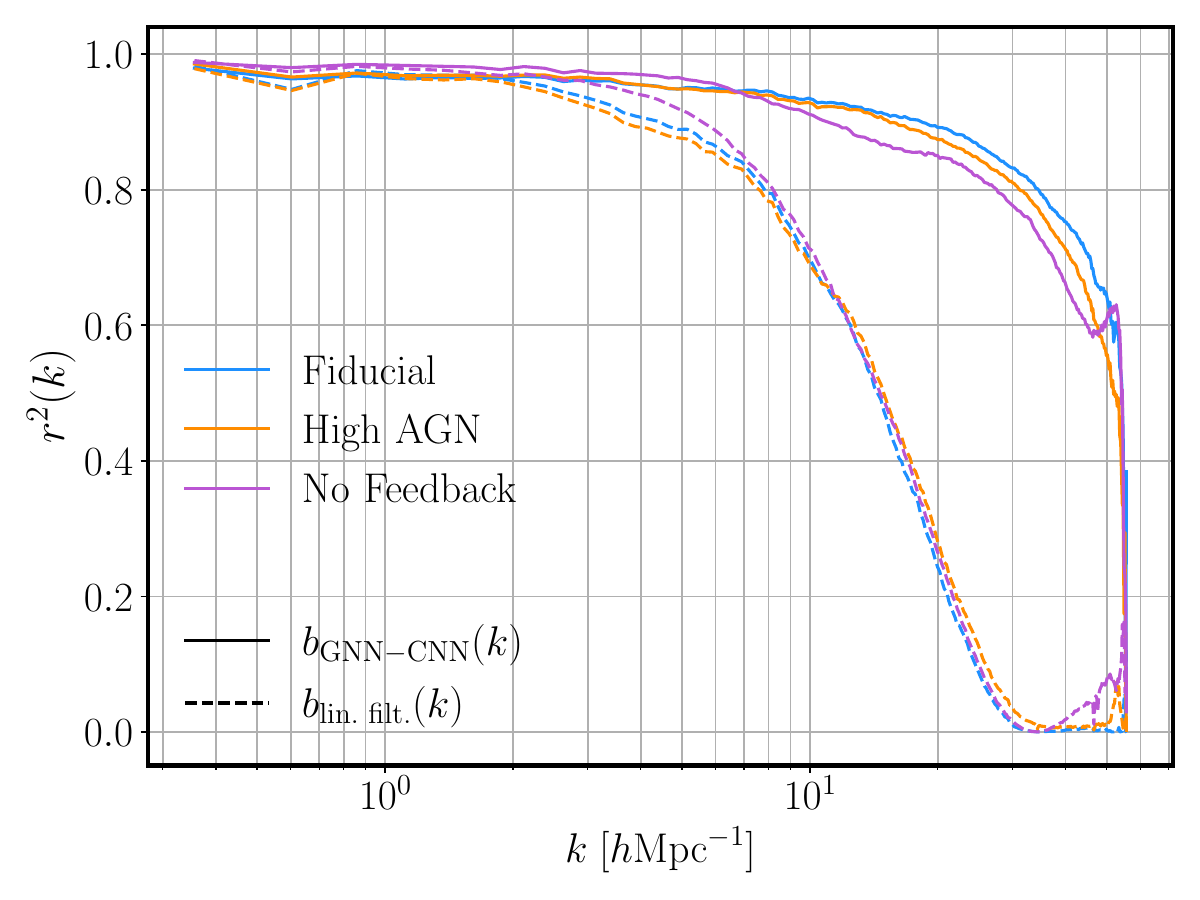}
  \caption{The cross-correlation coefficients of the truth with the GNN-CNN (straight) and linear transfer function (dashed) dark matter fields. The simulations are fiducial (blue), high AGN feedback (orange), and no feedback (purple).}
\label{fig:rk_ex_gnncnnlinfilt}
\end{figure}

For this test, we again only evaluate the performance of GNN-CNN and linear transfer function methods. The GNN-CNN is trained on a subset of the CV simulations. For the  linear transfer function, the `training' step involves applying a filter (Eq.~(\ref{eq:transferk})) derived from one of the CV simulations to the EX simulations. In Fig.~(\ref{fig:rk_ex_gnncnnlinfilt}), the cross-correlation coefficients for the GNN-CNN and linear filter fields change only minimally between the simulations. If there are errors with the individual halo profiles, these get diluted by averaging across all the halos in the simulation. So, even for the most extreme case where there is no feedback, the cross-correlation is not strongly impacted because there are so many halos.

\begin{figure}[h!]
\centering
  \includegraphics[width=0.7\linewidth]{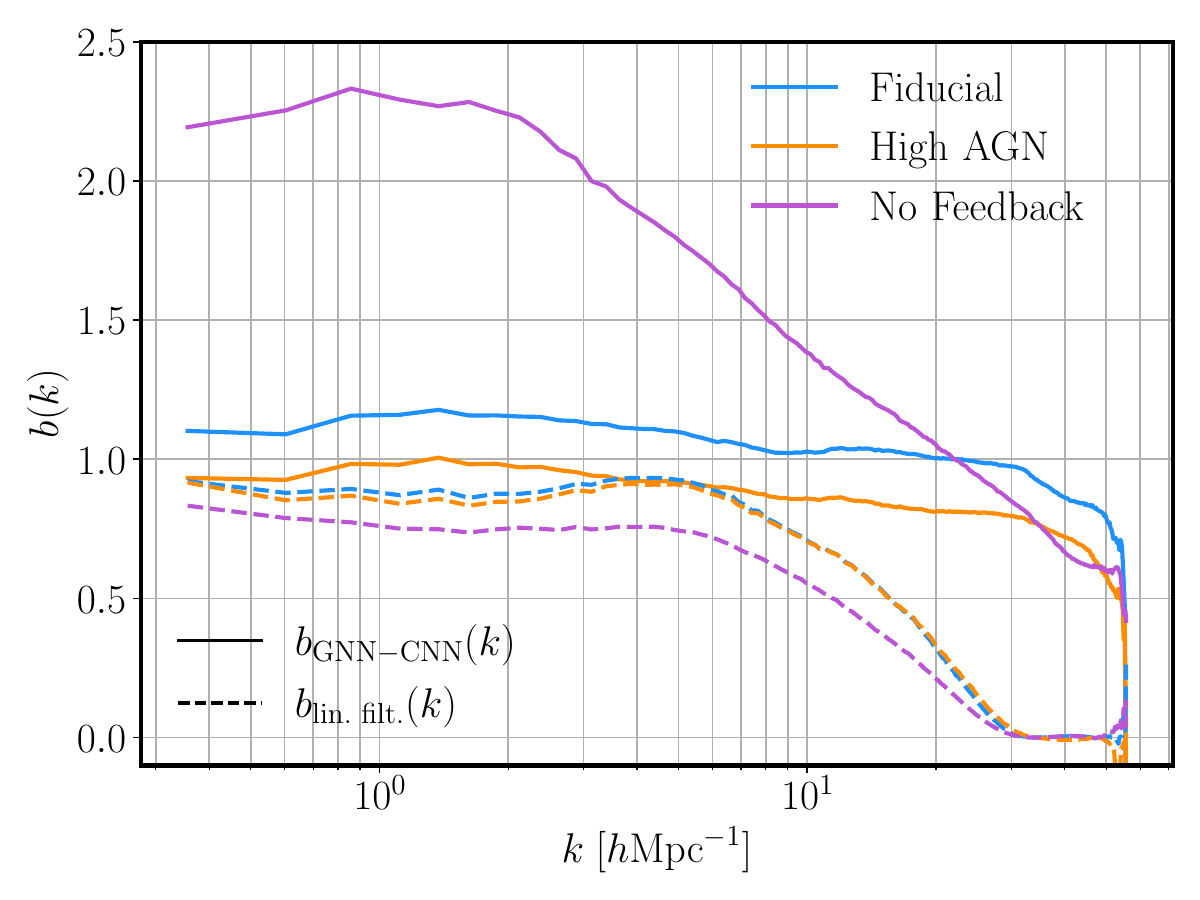}
  \caption{The bias of the reconstructed power spectra to the truth, for the GNN-CNN (straight) and  linear transfer function (dashed) reconstructions. The simulations are fiducial (blue), high AGN feedback (orange), and no feedback (purple). 
  }
  \label{fig:bk_ex_gnncnnlinfilt}
\end{figure}

To quantify how biased the results are, we plot $b(k)=P_{\rm{rec\,x\, true}} (k)/P_{\rm{true}}(k)$ where $P_{\rm{rec}}(k)$ is the GNN-CNN or  linear transfer function result. This is shown in Fig.~(\ref{fig:bk_ex_gnncnnlinfilt}). 
For both the GNN-CNN and linear transfer function methods, the bias generally remains scale-independent. The main exception is the GNN-CNN no feedback reconstruction, which also exhibits the largest bias. This significant drop in performance for the no feedback case is likely due to the extreme change in galaxy number density; this simulation contains over three times the number of galaxies present in the training set. It is promising that the very efficient AGN simulation does not result in a significant bias. Additionally, we anticipate that the GNN-CNN's performance could improve when applied to different simulations if trained on simulations with varying cosmological and astrophysical parameters. We briefly explored in our prior work, \cite{Kvasiuk:2024kwe}, but only assessed the cross-correlation coefficients not the bias.

\textbf{Robust cosmological constraints with squeezed limit observables.} One may ask whether a ML model trained on simulations, which will have incorrect baryonic physics compared to reality, can ever be used in practice for cosmological parameter analysis. A clean case where such baryonic uncertainty can be marginalized over, and one can thus achieve robust results, are squeezed limit observables. A recent example for using ML in a squeezed limit analysis is \cite{Giri:2022nzt,Kvasiuk:2024gbz}, where a neural network was used to estimate $f_{NL}$ in a way that is robust to baryonic uncertainty. This is possible because small-scale baryonic physics results in a constant multiplicative bias parameter on large scales that can be marginalized over when estimating cosmological parameters. Here, we are proposing a different application of ML in the squeezed limit. We can use our GNN-CNN method to reconstruct the electron density from observed galaxies, and then use this estimate as an input field to the quadratic kSZ velocity reconstruction estimator. As we recall in Appendix~(\ref{app:bv}), a mismatch between the true (in reality) and assumed (in simulations or theoretical modeling) electron distribution results in a multiplicative bias of the reconstructed velocity. This bias can easily be marginalized over when obtaining cosmological parameter constraints from kSZ velocity reconstruction. Nevertheless, as we showed in our previous work \cite{Kvasiuk:2023nje} and recall in Appendix~(\ref{app:bv}), a better electron reconstruction raises the overall signal-to-noise in the analysis, and thus tightens cosmological constraints. We can thus robustly apply our GNN-CNN method in this setup, and obtain unbiased cosmological constraints despite baryonic uncertainty in the training data of our GNN-CNN. This will be explored in more detail in future work. We note however that this robustness only holds for squeezed limit observables. In the next section we briefly comment on more general analyses.

\section{Outlook: Using the Model for Parameter Inference}\label{sec:paramdepinf}

The purpose of our work is to make a template of unobserved fields for cross-correlation analyses, such as gravitational lensing (dark matter template) or kSZ tomography (gas template). The observed cross-correlation will be strongest if the cosmological and astrophysical parameters assumed by the neural network model match those of reality. If we train our model as a function of cosmological and astrophysical parameters $\Lambda$ (a vector of parameters), we can thus use it for parameter inference. 

We outline how this works conceptually. Assume that we have trained the neural network 
\begin{eqnarray}
   \hat{\delta}^{NN}_m = \mathrm{NN}_\theta(\delta_g^{\mathrm{obs}}, \Lambda), 
\end{eqnarray}
from a data set that spans a range of these values $\Lambda$ using a MSE loss
\begin{eqnarray}
\mathcal{L}_{\mathrm{train}}(\theta) = \mathbb{E}_{(\delta_g^{\mathrm{obs}}, \Lambda, \delta_m) \sim \mathcal{D}} \left[ \left\| \mathrm{NN}_\theta(\delta_g^{\mathrm{obs}}, \Lambda) - \delta_m \right\|_2^2 \right].
\end{eqnarray}

We make the simplifying assumption that we have also measured the true dark matter density $\delta^{\mathrm{lens}}_m$ directly from gravitational lensing (in reality, reconstructing the 3D matter distribution from lensing can only be done approximately). We are thus given a pair \((\delta_g^{\mathrm{obs}}, \delta^\mathrm{lens}_m)\), but \(\Lambda\) in reality is unknown. We wish to estimate \(\Lambda\) such that the neural network prediction \(\hat{\delta}_m = \mathrm{NN}_\theta(\delta_g^{\mathrm{obs}}, \Lambda)\) matches the known \(\delta^\mathrm{lens}_m\). We do this by solving the following optimization problem:
\begin{eqnarray}
\hat{\Lambda} = \arg\min_{\Lambda} \; \mathcal{L}_{\mathrm{infer}}(\Lambda), \quad \text{where} \quad \mathcal{L}_{\mathrm{infer}}(\Lambda) = \left\| \mathrm{NN}_\theta(\delta_g^{\mathrm{obs}}, \Lambda) - \delta^\mathrm{lens}_m \right\|_2^2.
\end{eqnarray}
This can be done via gradient-based optimization (e.g., Adam) since our model is auto-differentiable. For simplicity, we have formulated the above optimization problem in terms of maximum likelihood, but it would be straight forward to generalize this to a Bayesian formulation including priors on the $\Lambda$ parameters.

To obtain uncertainty on the estimated parameters \(\hat{\Lambda}\), we use a Laplace approximation of the posterior \(p(\Lambda \mid \delta_g^{\mathrm{obs}}, \delta^\mathrm{lens}_m)\), assuming a locally Gaussian form around the optimum:
\begin{eqnarray}
p(\Lambda \mid \delta_g^{\mathrm{obs}}, \delta^\mathrm{lens}_m) \approx \mathcal{N}(\hat{\Lambda}, \Sigma_\Lambda).
\end{eqnarray}
The covariance matrix is obtained from the Hessian at the optimal point
\begin{eqnarray}
H = \nabla^2_{\Lambda} \mathcal{L}_{\mathrm{inv}}(\Lambda) \Big|_{\Lambda = \hat{\Lambda}} \quad \Rightarrow \quad \Sigma_\Lambda = H^{-1}
\end{eqnarray}
and provides the error bars on the parameters. For example, if a scalar parameter $\Lambda_i$ has a small error bar $\sigma_{\Lambda_i} = (H^{-1/2})_{ii}$, that means that this parameter is well determined by the data. Conversely, if $\sigma_{\Lambda_i}$ is large, it means that the reconstruction is not sensitive to this parameter (which can also be interpreted as the reconstruction being robust to this parameter). 

Of course, whether one can make interesting measurements in this way depends on whether the training data are sufficiently realistic, i.e. whether it models small-scale physics sufficiently accurately rather than being dominated by systematic simulation uncertainties. It is not likely that interesting cosmology constraints can be obtained in this way; however, for small-scale astrophysics parameters, the model should give some insights, for example whether AGN feedback is small or large. Contrary to other methods, the approach described here is not based on computationally expensive sampling of a forward model (with a probabilistic dark matter to galaxy connection) and may thus be easier to perform in practice. Since it works at field level, in principle it has access to all information to the extent that the Gaussian approximations above hold in practice. 

\section{Conclusions and Outlook}\label{sec:conc}

In this paper, we compared different methods of reconstructing continuous fields, in particular dark matter, from discrete tracers. We compared linear transfer function methods (which includes painting uniform profiles on each object), individual halo painting (depending on the halo mass), and finally field-level ML. As a baseline, we used a simple stellar mass-weighted grid assignment of galaxies. Our findings are:
\begin{itemize}
    \item Although the linear transfer function method has the same correlation coefficient with the true dark matter field as the unfiltered baseline (Fig.~(\ref{fig:rk_pk_dm_compareall})), it is a better representation at the field level  with smaller residuals (Fig.~(\ref{fig:all_model_otp_compare})). 
    \item We find that reconstructing dark matter by painting individual spherical halos with mass-dependent NFW profiles increases the cross-correlation with the true dark matter distribution \emph{within} the radius of the halo (Fig.~(\ref{fig:rk_dm_nfw_masknotall})), but lowers the cross-correlation with the entire true dark matter field (Fig.~(\ref{fig:rk_pk_dm_compareall})).
    \item While we find that a GNN can predict the mass of the host halo quite well (Fig.~(\ref{fig:GNN_preds})), this does not translate into an improved reconstruction of the dark matter field compared to simple mass assignment of galaxies.
    \item We find that the non-linearity and expressiveness of the field-level neural network is helpful in improving the reconstruction. The GNN-CNN approach clearly outperforms simple grid assignment or halo model approaches (Fig.~(\ref{fig:rk_pk_dm_compareall})). The cross-correlation coefficient shows improvement across all scales, notably extending to small-scales where, for instance, $r^2(k)\big|_{k=3\ \texttt{Mpc/h}}\approx 0.8$, a significant gain over other methods that had already dropped to zero. 
    \item The GNN-CNN was the most effective method for reconstructing the gas field (Fig.~(\ref{fig:rk_pk_gas_compareall})), underscoring the critical role of field-level methods given that gas largely resides outside of halos.
    \item We found the GNN-CNN to be fairly robust against astrophysical biases when applied to simulations with extremely different astrophysics than the training data (Fig.~(\ref{fig:rk_ex_gnncnnlinfilt})). The linear transfer function proved to be more robust to extrapolating to simulations with varying astrophysics. However, we could further improve the GNN-CNN performance by training on a simulation suite with varying astrophysical and cosmological parameters.  
\end{itemize}

Ultimately, the GNN-CNN ML approach delivered the best results, for dark matter and gas. This method correctly represents galaxies as point clouds and improves the reconstruction of small-scale details in the gridded field. The GNN-CNN notably improved the cross-correlation coefficient with the true field across all scales compared to our baseline, with the most significant improvements on small-scales. 

It would be interesting to train our neural network method on existing and upcoming high-resolution large-volume simulations such as FLAMINGO \cite{Schaye:2023jqv} and MilleniumTNG \cite{Pakmor:2022yyn}. Because they span a larger volume, one could construct a more realistic mock survey, including a survey mask. In this work, we chose to use CAMELS because this set includes simulations with different astrophysical and cosmological parameters. However, the small volume of CAMELS limits both the largest scales that we can probe and the range of cluster masses in the training data. To apply the method to real data, one would need to make a more realistic galaxy set, in redshift space and with galaxy features that include appropriate noise. Applying the method to a large survey volume would also require a convolutional or patch-wise implementation. We note, however, that no training simulation with the full survey volume is needed, since we aim to learn small-scale nonlinear physics only.

In summary, the GNN-CNN method showed a clear improvement over the linear filter technique, which is currently employed in applications such as large-scale velocity reconstruction with kSZ \cite{Deutsch:2017_2Dksz_estim1,Smith:2018bpn,Cayuso:2021ljq,McCarthy:2024_ACT_DESI_ksz,Bloch:2024PlanckxUnwise,Lague:2024czc_7.2sigma,Krywonos:2024mpb,Lai:2025qdw,Hotinli:2025tul} and searches for light bosons using resonant conversion \cite{mondino2024axioninducedpatchyscreeningcosmic, mccarthy2024darkphotonlimitspatchy,P_rvu_2024}. This indicates that adopting the GNN-CNN setup could increase the quality of these and similar analyses. We show this explicitly for kSZ velocity reconstruction in Appendix~(\ref{app:bv}), and also discuss how to marginalize over baryonic uncertainty in this case. More broadly, this technique's ability to provide a field-level reconstruction of dark matter — the fundamental scaffolding of the universe's matter content — makes it generally applicable to cross-correlation studies, with probes like cosmic shear, 21-cm intensity maps, fast radio bursts, and gravitational wave sources. We explore these and other applications in future work. 

\paragraph{Acknowledgments}

We thank H. Ganjoo for discussions and collaboration on alternative architectures for reconstructing cosmological fields. We thank C. K. Jespersen for discussions on graph neural networks. We are grateful to Francisco Villaescusa-Navarro for the useful clarification about the CAMELS dataset. M.M. acknowledges the support by the U.S. Department of Energy, Office of Science, Office of High Energy Physics under Award Number DE-SC-0017647, and by the National Science Foundation (NSF) under Grant Number 2307109. M.C.J. is supported by the National Science and Engineering Research Council through a Discovery grant. J.K. acknowledges support from the Natural Sciences and Engineering Research Council of Canada (NSERC) through the Vanier Canada Graduate Scholarship. Y.K. and M.M. are grateful for the hospitality of Perimeter Institute, where a part of this work was done. Research at Perimeter Institute is supported in part by the Government of Canada through the Department of Innovation, Science and Economic Development Canada and by the Province of Ontario through the Ministry of Colleges and Universities.

\addcontentsline{toc}{section}{References}
\bibliographystyle{JHEP}
\bibliography{References}

\providecommand{\href}[2]{#2}\begingroup\raggedright\begin{thebibliography}{10}

\bibitem{battaglia2018relationalinductivebiasesdeep}
P.W.~Battaglia, J.B.~Hamrick, V.~Bapst, A.~Sanchez-Gonzalez, V.~Zambaldi, M.~Malinowski et~al., \emph{Relational inductive biases, deep learning, and graph networks},  2018.

\bibitem{garuda2024estimatingdarkmatterhalo}
N.~Garuda, J.F.~Wu, D.~Nelson and A.~Pillepich, \emph{Estimating dark matter halo masses in simulated galaxy clusters with graph neural networks},  2024.

\bibitem{VD_hm}
P.~Villanueva-Domingo, F.~Villaescusa-Navarro, D.~Anglés-Alcázar, S.~Genel, F.~Marinacci, D.N.~Spergel et~al., \emph{Inferring halo masses with graph neural networks}, \href{https://doi.org/10.3847/1538-4357/ac7aa3}{\emph{The Astrophysical Journal} {\bfseries 935} }.

\bibitem{VD_siom}
P.~Villanueva-Domingo and F.~Villaescusa-Navarro, \emph{Learning cosmology and clustering with cosmic graphs}, \href{https://doi.org/10.3847/1538-4357/ac8930}{\emph{The Astrophysical Journal} {\bfseries 937} (2022) 115}.

\bibitem{ravanbakhsh2017estimatingcosmologicalparametersdark}
S.~Ravanbakhsh, J.~Oliva, S.~Fromenteau, L.C.~Price, S.~Ho, J.~Schneider et~al., \emph{Estimating cosmological parameters from the dark matter distribution},  2017.

\bibitem{schmelzle2017cosmologicalmodeldiscriminationdeep}
J.~Schmelzle, A.~Lucchi, T.~Kacprzak, A.~Amara, R.~Sgier, A.~Réfrégier et~al., \emph{Cosmological model discrimination with deep learning},  2017.

\bibitem{Gupta_2018}
A.~Gupta, J.M.Z.~Matilla, D.~Hsu and Z.~Haiman, \emph{Non-gaussian information from weak lensing data via deep learning}, \href{https://doi.org/10.1103/physrevd.97.103515}{\emph{Physical Review D} {\bfseries 97} (2018) }.

\bibitem{Ribli_2019}
D.~Ribli, B.A.~Pataki, J.M.~Zorrilla~Matilla, D.~Hsu, Z.~Haiman and I.~Csabai, \emph{Weak lensing cosmology with convolutional neural networks on noisy data}, \href{https://doi.org/10.1093/mnras/stz2610}{\emph{Monthly Notices of the Royal Astronomical Society} {\bfseries 490} (2019) 1843–1860}.

\bibitem{Fluri_2019}
J.~Fluri, T.~Kacprzak, A.~Lucchi, A.~Refregier, A.~Amara, T.~Hofmann et~al., \emph{Cosmological constraints with deep learning from kids-450 weak lensing maps}, \href{https://doi.org/10.1103/physrevd.100.063514}{\emph{Physical Review D} {\bfseries 100} (2019) }.

\bibitem{Matilla_2020}
J.M.Z.~Matilla, M.~Sharma, D.~Hsu and Z.~Haiman, \emph{Interpreting deep learning models for weak lensing}, \href{https://doi.org/10.1103/physrevd.102.123506}{\emph{Physical Review D} {\bfseries 102} (2020) }.

\bibitem{villaescusanavarro2021multifieldcosmologyartificialintelligence}
F.~Villaescusa-Navarro, D.~Anglés-Alcázar, S.~Genel, D.N.~Spergel, Y.~Li, B.~Wandelt et~al., \emph{Multifield cosmology with artificial intelligence},  2021.

\bibitem{villaescusanavarro2021robustmarginalizationbaryoniceffects}
F.~Villaescusa-Navarro, S.~Genel, D.~Angles-Alcazar, D.N.~Spergel, Y.~Li, B.~Wandelt et~al., \emph{Robust marginalization of baryonic effects for cosmological inference at the field level},  2021.

\bibitem{Villanueva_Domingo_2021}
P.~Villanueva-Domingo and F.~Villaescusa-Navarro, \emph{Removing astrophysics in 21 cm maps with neural networks}, \href{https://doi.org/10.3847/1538-4357/abd245}{\emph{The Astrophysical Journal} {\bfseries 907} (2021) 44}.

\bibitem{Lazanu_2021}
A.~Lazanu, \emph{Extracting cosmological parameters from n-body simulations using machine learning techniques}, \href{https://doi.org/10.1088/1475-7516/2021/09/039}{\emph{Journal of Cosmology and Astroparticle Physics} {\bfseries 2021} (2021) 039}.

\bibitem{Lu_2022}
T.~Lu, Z.~Haiman and J.M.~Zorrilla Matilla, \emph{Simultaneously constraining cosmology and baryonic physics via deep learning from weak lensing}, \href{https://doi.org/10.1093/mnras/stac161}{\emph{Monthly Notices of the Royal Astronomical Society} {\bfseries 511} (2022) 1518–1528}.

\bibitem{Kvasiuk:2024kwe}
Y.~Kvasiuk, J.~Krywonos, M.C.~Johnson and M.~M\"unchmeyer, \emph{{Reconstruction of Continuous Cosmological Fields from Discrete Tracers with Graph Neural Networks}},  in \emph{{38th conference on Neural Information Processing Systems}}, 11, 2024 [\href{https://arxiv.org/abs/2411.02496}{{\ttfamily 2411.02496}}].

\bibitem{CAMELS_presentation}
F.~{Villaescusa-Navarro}, D.~{Angl{\'e}s-Alc{\'a}zar}, S.~{Genel}, D.N.~{Spergel}, R.S.~{Somerville}, R.~{Dave} et~al., \emph{{The CAMELS Project: Cosmology and Astrophysics with Machine-learning Simulations}}, \href{https://doi.org/10.3847/1538-4357/abf7ba}{\emph{Astrophys. J.} {\bfseries 915} (2021) 71} [\href{https://arxiv.org/abs/2010.00619}{{\ttfamily 2010.00619}}].

\bibitem{CAMELS_DR1}
F.~{Villaescusa-Navarro}, S.~{Genel}, D.~{Angl{\'e}s-Alc{\'a}zar}, L.A.~{Perez}, P.~{Villanueva-Domingo}, D.~{Wadekar} et~al., \emph{{The CAMELS Project: Public Data Release}}, \href{https://doi.org/10.3847/1538-4365/acbf47}{\emph{Astrophys. J. Suppl.} {\bfseries 265} (2023) 54} [\href{https://arxiv.org/abs/2201.01300}{{\ttfamily 2201.01300}}].

\bibitem{CAMELS_DR2}
Y.~{Ni}, S.~{Genel}, D.~{Angl{\'e}s-Alc{\'a}zar}, F.~{Villaescusa-Navarro}, Y.~{Jo}, S.~{Bird} et~al., \emph{{The CAMELS Project: Expanding the Galaxy Formation Model Space with New ASTRID and 28-parameter TNG and SIMBA Suites}}, \href{https://doi.org/10.3847/1538-4357/ad022a}{\emph{Astrophys. J.} {\bfseries 959} (2023) 136} [\href{https://arxiv.org/abs/2304.02096}{{\ttfamily 2304.02096}}].

\bibitem{CMD}
F.~{Villaescusa-Navarro}, S.~{Genel}, D.~{Angles-Alcazar}, L.~{Thiele}, R.~{Dave}, D.~{Narayanan} et~al., \emph{{The CAMELS Multifield Dataset: Learning the Universe's Fundamental Parameters with Artificial Intelligence}}, {\emph{arXiv e-prints} (2021) arXiv:2109.10915} [\href{https://arxiv.org/abs/2109.10915}{{\ttfamily 2109.10915}}].

\bibitem{Sharma:2024kwj}
D.~Sharma, B.~Dai, F.~Villaescusa-Navarro and U.~Seljak, \emph{{A field-level emulator for modelling baryonic effects across hydrodynamic simulations}}, \href{https://doi.org/10.1093/mnras/staf355}{\emph{Mon. Not. Roy. Astron. Soc.} {\bfseries 538} (2025) 1415} [\href{https://arxiv.org/abs/2401.15891}{{\ttfamily 2401.15891}}].

\bibitem{Liu:2025gba}
R.H.~Liu, B.~Hadzhiyska, S.~Ferraro, S.~Bose and C.~Hern\'andez-Aguayo, \emph{{Fast Baryonic Field Painting for Sunyaev-Zel'dovich Analyses: Transfer Function vs. Hybrid Effective Field Theory}},  \href{https://arxiv.org/abs/2504.11794}{{\ttfamily 2504.11794}}.

\bibitem{Cooray:2002dia}
A.~Cooray and R.K.~Sheth, \emph{{Halo Models of Large Scale Structure}}, \href{https://doi.org/10.1016/S0370-1573(02)00276-4}{\emph{Phys. Rept.} {\bfseries 372} (2002) 1} [\href{https://arxiv.org/abs/astro-ph/0206508}{{\ttfamily astro-ph/0206508}}].

\bibitem{Asgari_2023}
M.~Asgari, A.J.~Mead and C.~Heymans, \emph{The halo model for cosmology: a pedagogical review}, \href{https://doi.org/10.21105/astro.2303.08752}{\emph{The Open Journal of Astrophysics} {\bfseries 6} (2023) }.

\bibitem{Wechsler_2018}
R.H.~Wechsler and J.L.~Tinker, \emph{The connection between galaxies and their dark matter halos}, \href{https://doi.org/10.1146/annurev-astro-081817-051756}{\emph{Annual Review of Astronomy and Astrophysics} {\bfseries 56} (2018) 435–487}.

\bibitem{Zhai:2024knl}
Z.~Zhai and W.J.~Percival, \emph{{Testing the framework of the halo occupation distribution with assembly bias modelling and empirical extensions}}, \href{https://doi.org/10.1093/mnras/stae2489}{\emph{Mon. Not. Roy. Astron. Soc.} {\bfseries 535} (2024) 2469} [\href{https://arxiv.org/abs/2409.19399}{{\ttfamily 2409.19399}}].

\bibitem{oppenheimer2025introducingdescriptiveparametricmodel}
B.D.~Oppenheimer, G.M.~Voit, Y.M.~Bahé, N.~Battaglia, J.~Bregman, J.N.~Burchett et~al., \emph{Introducing the descriptive parametric model: Gaseous profiles for galaxies, groups, and clusters},  2025.

\bibitem{Ginzburg_2017}
D.~Ginzburg, V.~Desjacques and K.C.~Chan, \emph{Shot noise and biased tracers: A new look at the halo model}, \href{https://doi.org/10.1103/physrevd.96.083528}{\emph{Physical Review D} {\bfseries 96} (2017) }.

\bibitem{Chen:2019wik}
A.Y.~Chen and N.~Afshordi, \emph{{Amending the halo model to satisfy cosmological conservation laws}}, \href{https://doi.org/10.1103/PhysRevD.101.103522}{\emph{Phys. Rev. D} {\bfseries 101} (2020) 103522} [\href{https://arxiv.org/abs/1912.04872}{{\ttfamily 1912.04872}}].

\bibitem{Sehgal_2010}
N.~Sehgal, P.~Bode, S.~Das, C.~Hernandez-Monteagudo, K.~Huffenberger, Y.-T.~Lin et~al., \emph{Simulations of the microwave sky}, \href{https://doi.org/10.1088/0004-637x/709/2/920}{\emph{The Astrophysical Journal} {\bfseries 709} (2010) 920–936}.

\bibitem{Stein_2018}
G.~Stein, M.A.~Alvarez and J.R.~Bond, \emph{The mass-peak patch algorithm for fast generation of deep all-sky dark matter halo catalogues and itsn-body validation}, \href{https://doi.org/10.1093/mnras/sty3226}{\emph{Monthly Notices of the Royal Astronomical Society} {\bfseries 483} (2018) 2236–2250}.

\bibitem{Stein_2020}
G.~Stein, M.A.~Alvarez, J.R.~Bond, A.v.~Engelen and N.~Battaglia, \emph{The websky extragalactic cmb simulations}, \href{https://doi.org/10.1088/1475-7516/2020/10/012}{\emph{Journal of Cosmology and Astroparticle Physics} {\bfseries 2020} (2020) 012–012}.

\bibitem{Yasini2020}
S.~Yasini, M.~Alvarez, E.~Schaan, K.~Maamari, S.K.~s.~Mazinani, N.~Mirzatuny et~al., \emph{Astropaint: A python package for painting halo catalogs into celestial maps}, \href{https://doi.org/10.21105/joss.02608}{\emph{Journal of Open Source Software} {\bfseries 5} (2020) 2608}.

\bibitem{InferringGalDH}
R.~von Marttens, L.~Casarini, N.R.~Napolitano, S.~Wu, V.~Amaro, R.~Li et~al., \emph{Inferring galaxy dark halo properties from visible matter with machine learning}, \href{https://doi.org/10.1093/mnras/stac2449}{\emph{Monthly Notices of the Royal Astronomical Society} {\bfseries 516} (2022) 3924}.

\bibitem{Villanueva-Domingo_2022}
P.~Villanueva-Domingo, F.~Villaescusa-Navarro, D.~Anglés-Alcázar, S.~Genel, F.~Marinacci, D.N.~Spergel et~al., \emph{Inferring halo masses with graph neural networks}, \href{https://doi.org/10.3847/1538-4357/ac7aa3}{\emph{The Astrophysical Journal} {\bfseries 935} (2022) 30}.

\bibitem{Calderon:2019}
V.F.~Calderon and A.A.~Berlind, \emph{{Prediction of galaxy halo masses in SDSS DR7 via a machine learning approach}}, \href{https://doi.org/10.1093/mnras/stz2775}{\emph{Mon. Not. Roy. Astron. Soc.} {\bfseries 490} (2019) 2367} [\href{https://arxiv.org/abs/1902.02680}{{\ttfamily 1902.02680}}].

\bibitem{2023MNRAS.525.6015D}
M.~{de los Rios}, M.~{Peta{\v{c}}}, B.~{Zaldivar}, N.R.~{Bonaventura}, F.~{Calore} and F.~{Iocco}, \emph{{Determining the dark matter distribution in simulated galaxies with deep learning}}, \href{https://doi.org/10.1093/mnras/stad2614}{\emph{"Monthly Notices of the Royal Astronomical Society"} {\bfseries 525} (2023) 6015} [\href{https://arxiv.org/abs/2111.08725}{{\ttfamily 2111.08725}}].

\bibitem{Hahn_2024}
C.~Hahn, C.~Bottrell and K.-G.~Lee, \emph{Haloflow. i. neural inference of halo mass from galaxy photometry and morphology}, \href{https://doi.org/10.3847/1538-4357/ad4344}{\emph{The Astrophysical Journal} {\bfseries 968} (2024) 90}.

\bibitem{2024A&A...686A..80W}
S.~{Wu}, N.R.~{Napolitano}, C.~{Tortora}, R.~{von Marttens}, L.~{Casarini}, R.~{Li} et~al., \emph{{Total and dark mass from observations of galaxy centers with machine learning}}, \href{https://doi.org/10.1051/0004-6361/202348152}{\emph{Astronomy \& Astrophysics} {\bfseries 686} (2024) A80} [\href{https://arxiv.org/abs/2310.02816}{{\ttfamily 2310.02816}}].

\bibitem{ono2024debiasingdiffusionprobabilisticreconstruction}
V.~Ono, C.F.~Park, N.~Mudur, Y.~Ni, C.~Cuesta-Lazaro and F.~Villaescusa-Navarro, \emph{Debiasing with diffusion: Probabilistic reconstruction of dark matter fields from galaxies with camels},  2024.

\bibitem{park2023probabilisticreconstructiondarkmatter}
C.F.~Park, V.~Ono, N.~Mudur, Y.~Ni and C.~Cuesta-Lazaro, \emph{Probabilistic reconstruction of dark matter fields from biased tracers using diffusion models},  2023.

\bibitem{Wang:2023hgm}
Z.~Wang, F.~Shi, X.~Yang, Q.~Li, Y.~Liu and X.~Li, \emph{{(DarkAI) Mapping the large-scale density field of dark matter using artificial intelligence}}, \href{https://doi.org/10.1007/s11433-023-2192-9}{\emph{Sci. China Phys. Mech. Astron.} {\bfseries 67} (2024) 219513} [\href{https://arxiv.org/abs/2305.11431}{{\ttfamily 2305.11431}}].

\bibitem{Shi:2025zoz}
F.~Shi et~al., \emph{{DarkAI: Reconstructing the Density, Velocity, and Tidal Fields of Dark Matter from a DESI-like Bright Galaxy Sample}}, \href{https://doi.org/10.3847/1538-4365/adfa26}{\emph{Astrophys. J. Suppl.} {\bfseries 280} (2025) 53} [\href{https://arxiv.org/abs/2501.12621}{{\ttfamily 2501.12621}}].

\bibitem{Jasche_2013}
J.~Jasche and B.D.~Wandelt, \emph{Bayesian physical reconstruction of initial conditions from large-scale structure surveys}, \href{https://doi.org/10.1093/mnras/stt449}{\emph{Monthly Notices of the Royal Astronomical Society} {\bfseries 432} (2013) 894–913}.

\bibitem{2013ApJ...779...15J}
J.~{Jasche} and B.D.~{Wandelt}, \emph{{Methods for Bayesian Power Spectrum Inference with Galaxy Surveys}}, \href{https://doi.org/10.1088/0004-637X/779/1/15}{\emph{"Astrophys. J."} {\bfseries 779} (2013) 15} [\href{https://arxiv.org/abs/1306.1821}{{\ttfamily 1306.1821}}].

\bibitem{Jasche_2014}
J.~Jasche and G.~Lavaux, \emph{Matrix-free large-scale bayesian inference in cosmology}, \href{https://doi.org/10.1093/mnras/stu2479}{\emph{Monthly Notices of the Royal Astronomical Society} {\bfseries 447} (2014) 1204–1212}.

\bibitem{2016MNRAS.455.3169L}
G.~{Lavaux} and J.~{Jasche}, \emph{{Unmasking the masked Universe: the 2M++ catalogue through Bayesian eyes}}, \href{https://doi.org/10.1093/mnras/stv2499}{\emph{"Mon. Not. Roy. Astron. Soc."} {\bfseries 455} (2016) 3169} [\href{https://arxiv.org/abs/1509.05040}{{\ttfamily 1509.05040}}].

\bibitem{Modi:2018cfi}
C.~Modi, Y.~Feng and U.~Seljak, \emph{{Cosmological Reconstruction From Galaxy Light: Neural Network Based Light-Matter Connection}}, \href{https://doi.org/10.1088/1475-7516/2018/10/028}{\emph{JCAP} {\bfseries 10} (2018) 028} [\href{https://arxiv.org/abs/1805.02247}{{\ttfamily 1805.02247}}].

\bibitem{Horowitz:2022fvl}
B.~Horowitz, C.~Hahn, F.~Lanusse, C.~Modi and S.~Ferraro, \emph{{Differentiable stochastic halo occupation distribution}}, \href{https://doi.org/10.1093/mnras/stae350}{\emph{Mon. Not. Roy. Astron. Soc.} {\bfseries 529} (2024) 2473} [\href{https://arxiv.org/abs/2211.03852}{{\ttfamily 2211.03852}}].

\bibitem{Wu_2024}
S.~Wu, N.R.~Napolitano, C.~Tortora, R.~von Marttens, L.~Casarini, R.~Li et~al., \emph{Total and dark mass from observations of galaxy centers with machine learning}, \href{https://doi.org/10.1051/0004-6361/202348152}{\emph{Astronomy \& Astrophysics} {\bfseries 686} (2024) A80}.

\bibitem{M_nchmeyer_2019}
M.~Münchmeyer, M.S.~Madhavacheril, S.~Ferraro, M.C.~Johnson and K.M.~Smith, \emph{Constraining local non-gaussianities with kinetic sunyaev-zel’dovich tomography}, \href{https://doi.org/10.1103/physrevd.100.083508}{\emph{Physical Review D} {\bfseries 100} (2019) }.

\bibitem{Hadzhiyska_2023}
B.~Hadzhiyska, L.~Hernquist, D.~Eisenstein, A.M.~Delgado, S.~Bose, R.~Kannan et~al., \emph{The millenniumtng project: refining the one-halo model of red and blue galaxies at different redshifts}, \href{https://doi.org/10.1093/mnras/stad279}{\emph{Monthly Notices of the Royal Astronomical Society} {\bfseries 524} (2023) 2524–2538}.

\bibitem{Pylians}
F.~{Villaescusa-Navarro}, ``{Pylians: Python libraries for the analysis of numerical simulations}.'' Astrophysics Source Code Library, record ascl:1811.008, Nov., 2018.

\bibitem{Child_2018}
H.L.~Child, S.~Habib, K.~Heitmann, N.~Frontiere, H.~Finkel, A.~Pope et~al., \emph{Halo profiles and the concentration–mass relation for a {$\Lambda$CDM} universe}, \href{https://doi.org/10.3847/1538-4357/aabf95}{\emph{The Astrophysical Journal} {\bfseries 859} (2018) 55}.

\bibitem{wang2019edgenet}
Y.~Wang, Y.~Sun, Z.~Liu, S.E.~Sarma, M.M.~Bronstein and J.M.~Solomon, \emph{Dynamic graph cnn for learning on point clouds}, {\emph{ACM Transactions on Graphics (tog)} {\bfseries 38} (2019) 1}.

\bibitem{Ho:2009iw}
S.~Ho, S.~Dedeo and D.~Spergel, \emph{{Finding the Missing Baryons Using CMB as a Backlight}},  \href{https://arxiv.org/abs/0903.2845}{{\ttfamily 0903.2845}}.

\bibitem{Zhang_2010}
P.~Zhang, \emph{The dark flow induced small-scale kinetic sunyaev--zel'dovich effect}, \href{https://doi.org/10.1111/j.1745-3933.2010.00899.x}{\emph{Monthly Notices of the Royal Astronomical Society: Letters} {\bfseries 407} (2010) L36}.

\bibitem{Shao_2011}
J.~Shao, P.~Zhang, W.~Lin, Y.~Jing and J.~Pan, \emph{Kinetic sunyaev-zel’dovich tomography with spectroscopic redshift surveys: Kinetic sz tomography}, \href{https://doi.org/10.1111/j.1365-2966.2011.18166.x}{\emph{Monthly Notices of the Royal Astronomical Society} {\bfseries 413} (2011) 628–642}.

\bibitem{Munshi:2015anr}
D.~Munshi, I.T.~Iliev, K.L.~Dixon and P.~Coles, \emph{{Extracting the late-time kinetic Sunyaev\textendash{}Zel'dovich effect}}, \href{https://doi.org/10.1093/mnras/stw2067}{\emph{Mon. Not. Roy. Astron. Soc.} {\bfseries 463} (2016) 2425} [\href{https://arxiv.org/abs/1511.03449}{{\ttfamily 1511.03449}}].

\bibitem{Deutsch:2017_2Dksz_estim1}
A.-S.~Deutsch, E.~Dimastrogiovanni, M.C.~Johnson, M.~M\"unchmeyer and A.~Terrana, \emph{{Reconstruction of the remote dipole and quadrupole fields from the kinetic Sunyaev Zel\textquoteright{}dovich and polarized Sunyaev Zel\textquoteright{}dovich effects}}, \href{https://doi.org/10.1103/PhysRevD.98.123501}{\emph{Phys. Rev. D} {\bfseries 98} (2018) 123501} [\href{https://arxiv.org/abs/1707.08129}{{\ttfamily 1707.08129}}].

\bibitem{Smith:2018bpn}
K.M.~Smith, M.S.~Madhavacheril, M.~M\"unchmeyer, S.~Ferraro, U.~Giri and M.C.~Johnson, \emph{{KSZ tomography and the bispectrum}},  \href{https://arxiv.org/abs/1810.13423}{{\ttfamily 1810.13423}}.

\bibitem{Cayuso:2021ljq}
J.~Cayuso, R.~Bloch, S.C.~Hotinli, M.C.~Johnson and F.~McCarthy, \emph{{Velocity reconstruction with the cosmic microwave background and galaxy surveys}}, \href{https://doi.org/10.1088/1475-7516/2023/02/051}{\emph{JCAP} {\bfseries 02} (2023) 051} [\href{https://arxiv.org/abs/2111.11526}{{\ttfamily 2111.11526}}].

\bibitem{McCarthy:2024_ACT_DESI_ksz}
F.~McCarthy et~al., \emph{{The Atacama Cosmology Telescope: Large-scale velocity reconstruction with the kinematic Sunyaev--Zel'dovich effect and DESI LRGs}},  \href{https://arxiv.org/abs/2410.06229}{{\ttfamily 2410.06229}}.

\bibitem{Bloch:2024PlanckxUnwise}
R.~Bloch and M.C.~Johnson, \emph{{Kinetic Sunyaev Zel'dovich velocity reconstruction from Planck and unWISE}},  \href{https://arxiv.org/abs/2405.00809}{{\ttfamily 2405.00809}}.

\bibitem{Lague:2024czc_7.2sigma}
A.~Lagu\"e, M.S.~Madhavacheril, K.M.~Smith, S.~Ferraro and E.~Schaan, \emph{{Constraints on local primordial non-Gaussianity with 3d Velocity Reconstruction from the Kinetic Sunyaev-Zeldovich Effect}},  \href{https://arxiv.org/abs/2411.08240}{{\ttfamily 2411.08240}}.

\bibitem{Krywonos:2024mpb}
J.~Krywonos, S.C.~Hotinli and M.C.~Johnson, \emph{{Constraints on cosmology beyond $\Lambda$CDM with kinetic Sunyaev Zel'dovich velocity reconstruction}},  \href{https://arxiv.org/abs/2408.05264}{{\ttfamily 2408.05264}}.

\bibitem{Lai:2025qdw}
A.C.M.~Lai, Y.~Kvasiuk and M.~M{\"u}nchmeyer, \emph{{KSZ Velocity Reconstruction with ACT and DESI-LS using a Tomographic QML Power Spectrum Estimator}},  \href{https://arxiv.org/abs/2506.21684}{{\ttfamily 2506.21684}}.

\bibitem{Hotinli:2025tul}
S.C.~Hotinli, K.M.~Smith and S.~Ferraro, \emph{{Velocity Reconstruction from KSZ: Measuring $f_{NL}$ with ACT and DESILS}},  \href{https://arxiv.org/abs/2506.21657}{{\ttfamily 2506.21657}}.

\bibitem{Sunyaev1980}
R.A.~Sunyaev and Y.B.~Zeldovich, \emph{{The velocity of clusters of galaxies relative to the microwave background. The possibility of its measurement}}, \href{https://doi.org/10.1093/mnras/190.3.413}{\emph{Monthly Notices of the Royal Astronomical Society} {\bfseries 190} (1980) 413}.

\bibitem{Kvasiuk:2023nje}
Y.~Kvasiuk and M.~M\"unchmeyer, \emph{{Autodifferentiable likelihood pipeline for the cross-correlation of CMB and large-scale structure due to the kinetic Sunyaev-Zeldovich effect}}, \href{https://doi.org/10.1103/PhysRevD.109.083515}{\emph{Phys. Rev. D} {\bfseries 109} (2024) 083515} [\href{https://arxiv.org/abs/2305.08903}{{\ttfamily 2305.08903}}].

\bibitem{Battaglia:2016xbi}
N.~Battaglia, \emph{{The Tau of Galaxy Clusters}}, \href{https://doi.org/10.1088/1475-7516/2016/08/058}{\emph{JCAP} {\bfseries 08} (2016) 058} [\href{https://arxiv.org/abs/1607.02442}{{\ttfamily 1607.02442}}].

\bibitem{Madhavacheril:2019buy}
M.S.~Madhavacheril, N.~Battaglia, K.M.~Smith and J.L.~Sievers, \emph{{Cosmology with the kinematic Sunyaev-Zeldovich effect: Breaking the optical depth degeneracy with fast radio bursts}}, \href{https://doi.org/10.1103/PhysRevD.100.103532}{\emph{Phys. Rev. D} {\bfseries 100} (2019) 103532} [\href{https://arxiv.org/abs/1901.02418}{{\ttfamily 1901.02418}}].

\bibitem{Giri:2020pkk}
U.~Giri and K.M.~Smith, \emph{{Exploring KSZ velocity reconstruction with N-body simulations and the halo~model}}, \href{https://doi.org/10.1088/1475-7516/2022/09/028}{\emph{JCAP} {\bfseries 09} (2022) 028} [\href{https://arxiv.org/abs/2010.07193}{{\ttfamily 2010.07193}}].

\bibitem{10.1093/mnras/stae2100}
L.~Bigwood, A.~Amon, A.~Schneider, J.~Salcido, I.G.~McCarthy, C.~Preston et~al., \emph{Weak lensing combined with the kinetic sunyaev–zel’dovich effect: a study of baryonic feedback}, \href{https://doi.org/10.1093/mnras/stae2100}{\emph{Monthly Notices of the Royal Astronomical Society} {\bfseries 534} (2024) 655} [\href{https://arxiv.org/abs/https://academic.oup.com/mnras/article-pdf/534/1/655/59261644/stae2100.pdf}{{\ttfamily https://academic.oup.com/mnras/article-pdf/534/1/655/59261644/stae2100.pdf}}].

\bibitem{Giri:2022nzt}
U.~Giri, M.~M{\"u}nchmeyer and K.M.~Smith, \emph{{Robust neural network-enhanced estimation of local primordial non-Gaussianity}}, \href{https://doi.org/10.1103/PhysRevD.107.L061301}{\emph{Phys. Rev. D} {\bfseries 107} (2023) L061301} [\href{https://arxiv.org/abs/2205.12964}{{\ttfamily 2205.12964}}].

\bibitem{Kvasiuk:2024gbz}
Y.~Kvasiuk, M.~M{\"u}nchmeyer and K.~Smith, \emph{{Two-field formalism for a neural network-enhanced non-Gaussianity search with halos}}, \href{https://doi.org/10.1103/2szy-wypg}{\emph{Phys. Rev. D} {\bfseries 112} (2025) 023540} [\href{https://arxiv.org/abs/2410.01007}{{\ttfamily 2410.01007}}].

\bibitem{Schaye:2023jqv}
J.~Schaye et~al., \emph{{The FLAMINGO project: cosmological hydrodynamical simulations for large-scale structure and galaxy cluster surveys}}, \href{https://doi.org/10.1093/mnras/stad2419}{\emph{Mon. Not. Roy. Astron. Soc.} {\bfseries 526} (2023) 4978} [\href{https://arxiv.org/abs/2306.04024}{{\ttfamily 2306.04024}}].

\bibitem{Pakmor:2022yyn}
R.~Pakmor et~al., \emph{{The MillenniumTNG Project: the hydrodynamical full physics simulation and a first look at its galaxy clusters}}, \href{https://doi.org/10.1093/mnras/stac3620}{\emph{Mon. Not. Roy. Astron. Soc.} {\bfseries 524} (2023) 2539} [\href{https://arxiv.org/abs/2210.10060}{{\ttfamily 2210.10060}}].

\bibitem{mondino2024axioninducedpatchyscreeningcosmic}
C.~Mondino, D.~Pîrvu, J.~Huang and M.C.~Johnson, \emph{Axion-induced patchy screening of the cosmic microwave background},  2024.

\bibitem{mccarthy2024darkphotonlimitspatchy}
F.~McCarthy, D.~Pirvu, J.C.~Hill, J.~Huang, M.C.~Johnson and K.K.~Rogers, \emph{Dark photon limits from patchy dark screening of the cosmic microwave background},  2024.

\bibitem{P_rvu_2024}
D.~Pîrvu, J.~Huang and M.C.~Johnson, \emph{Patchy screening of the cmb from dark photons}, \href{https://doi.org/10.1088/1475-7516/2024/01/019}{\emph{Journal of Cosmology and Astroparticle Physics} {\bfseries 2024} (2024) 019}.

\bibitem{Tishue:2025cvp}
A.J.~Tishue, C.~Shiveshwarkar and G.~Holder, \emph{{The kSZ optical depth degeneracy and future constraints on local primordial non-Gaussianity}},  \href{https://arxiv.org/abs/2510.25821}{{\ttfamily 2510.25821}}.

\bibitem{McCarthy:2025brx}
F.~McCarthy et~al., \emph{{The Atacama Cosmology Telescope: Cross-correlation of kSZ and continuity equation velocity reconstruction with photometric DESI LRGs}},  \href{https://arxiv.org/abs/2511.15701}{{\ttfamily 2511.15701}}.

\end{thebibliography}\endgroup

\appendix
\section{Marginalizing Over Baryonic Feedback in the Example of KSZ Velocity Reconstruction}
\label{app:bv}

We give an example of how our method can be used in practice, and how one can deal with baryonic uncertainty and obtain unbiased cosmology constraints in squeezed limit analyses. We show this explicitly in the case of the kSZ velocity reconstruction \cite{Deutsch:2017_2Dksz_estim1,Smith:2018bpn,Cayuso:2021ljq,McCarthy:2024_ACT_DESI_ksz,Bloch:2024PlanckxUnwise,Lague:2024czc_7.2sigma,Krywonos:2024mpb,Lai:2025qdw,Hotinli:2025tul}. Here, mismodelling of the electron power will result only in a multiplicative constant bias (summarizing and interpreting the results from \cite{Smith:2018bpn}). We also show why a better reconstruction of the electron field given observed galaxies leads to higher signal-to-noise. We start from the \textit{traditional} (ML free) quadratic estimator (QE) for the large-scale radial velocity mode $v_r(\mathbf{k}_L)$,
\begin{align}
    \hat v^{\rm trad}_r(\mathbf{k}_L)
    &=
    N_{v_r}(k_L)\,\frac{K_*}{\chi_*^2}
    \int \frac{d^3 k_S}{(2\pi)^3}
    \int \frac{d^2 \boldsymbol{\ell}}{(2\pi)^2}\,
    \frac{P_{ge}^{\rm fid}(k_S)}{P^{\rm tot}_{gg}(k_S)\,C_{\ell}^{TT,{\rm tot}}}\,
    \delta^{*\rm obs}_g(\mathbf{k}_S)\,T^*(\boldsymbol{\ell})\,
    (2\pi)^3\delta^{(3)}\!\left(\mathbf{k}_L+\mathbf{k}_S+\frac{\boldsymbol{\ell}}{\chi_*}\right),
    \label{eq:vr_estimator}
\end{align}
with normalization
\begin{align}
    N_{v_r}(k_L)
    &=
    \frac{\chi_*^4}{K_*^2}
    \left[
    \int \frac{d^3 k_S}{(2\pi)^3}
    \int \frac{d^2 \boldsymbol{\ell}}{(2\pi)^2}\,
    \frac{\big(P_{ge}^{\rm fid}(k_S)\big)^2}{P^{\rm tot}_{gg}(k_S)\,C_{\ell}^{TT,{\rm tot}}}\,
    (2\pi)^3\delta^{(3)}\!\left(\mathbf{k}_L+\mathbf{k}_S+\frac{\boldsymbol{\ell}}{\chi_*}\right)
    \right]^{-1}
    \nonumber\\
    &=
    \frac{\chi_*^2}{K_*^2}
    \left[
    \int \frac{k_S\,dk_S}{2\pi}\,
    \frac{\big(P_{ge}^{\rm fid}(k_S)\big)^2}{P^{\rm tot}_{gg}(k_S)\,C_{\ell}^{TT,{\rm tot}}}
    \Bigg|_{\ell = k_S\chi_*}
    \right]^{-1},
    \label{eq:vr_norm}
\end{align}
using the conventions of \cite{Smith:2018bpn}. The normalization enforces the correct expectation value 
\begin{align}
    \left< \hat v^{\rm trad}_r(\mathbf{k}_L) \right> = v_r(\mathbf{k}_L).
\end{align}

In the traditional kSZ estimator, we use $\delta_g$ as a tracer of the electron distribution because it is a directly observed quantity. We claim here that instead of $\delta^{\rm obs}_g$, we can use any \textit{local} function of $\delta^{\rm obs}_g$, such as the learned electron template $\hat{\delta}_e[\delta^{\rm obs}_g]$ (represented by the GNN-CNN above). In that case, the new (neural network-enhanced) QE is 
\begin{align}
    \hat v_r(\mathbf{k}_L)
    &=
    N_{v_r}(k_L)\,\frac{K_*}{\chi_*^2}
    \int \frac{d^3 k_S}{(2\pi)^3}
    \int \frac{d^2 \boldsymbol{\ell}}{(2\pi)^2}\,
    \frac{P_{\hat{e}e}^{\rm fid}(k_S)}{P^{\rm tot}_{\hat{e}\hat{e}}(k_S)\,C_{\ell}^{TT,{\rm tot}}}\,
    \hat{\delta}_e[\delta^{\rm obs}_g]^*(\mathbf{k}_S)\,T^*(\boldsymbol{\ell})\,
    (2\pi)^3\delta^{(3)}\!\left(\mathbf{k}_L+\mathbf{k}_S+\frac{\boldsymbol{\ell}}{\chi_*}\right),
    \label{eq:vr_estimator}
\end{align}
with normalization
\begin{align}
    N_{v_r}(k_L)
    =
    \frac{\chi_*^2}{K_*^2}
    \left[
    \int \frac{k_S\,dk_S}{2\pi}\,
    \frac{\big(P_{\hat{e}e}^{\rm fid}(k_S)\big)^2}{P^{\rm tot}_{\hat{g}\hat{g}}(k_S)\,C_{\ell}^{TT,{\rm tot}}}
    \Bigg|_{\ell = k_S\chi_*}
    \right]^{-1},
    \label{eq:vr_norm}
\end{align}
so that again
\begin{align}
    \left< \hat v_r(\mathbf{k}_L) \right> = v_r(\mathbf{k}_L).
\end{align}
As in the traditional case, we have to assume some $P^{\rm fid}_{\hat{e}e}(k_S) = \langle\hat{\delta}_e[\delta^{\rm obs}_g](\mathbf{k}_S)\delta_e(\mathbf{k}_S)\rangle$ - the correlation function of the small-scale estimator and the true field. We note that a choice of a transfer function $\hat{\delta}_e[\delta^{\rm obs}_g] = (P^{\rm fid}_{ge}(k_S)/P^{\rm tot}_{gg}(k_S))\delta^{\rm obs}_g(\mathbf{k}_S)$ gives exactly the traditional kSZ estimator. 

\subsection{Increasing the Signal-to-Noise}
 The noise is given by \eqref{eq:vr_norm}. We can rewrite it into the inverse noise as follows:
\begin{align}
    N_{v_r}^{-1}(k_L)
    &=
    \frac{K_*^2}{\chi_*^2}
    \int \frac{k_S\,dk_S}{2\pi}\,
    \frac{\big(P_{\hat{e}e}^{\rm fid}(k_S)\big)^2}
         {P^{\rm tot}_{\hat{e}\hat{e}}(k_S)\,C_{\ell}^{TT,{\rm tot}}}
    \Bigg|_{\ell=k_S\chi_*} \\
    &=  \frac{K_*^2}{\chi_*^2}\int \frac{k_S\,dk_S}{2\pi}\,\frac{P^{\rm fid}_{ee}(k_S)}{C^{TT,tot}_\ell}r_{e\hat{e}}^2(k_S),
\end{align}
where $r^2_{e\hat{e}}(k_{S})$ is a squared cross-correlation coefficient of the galaxy and electron field
\begin{equation}
    r_{e\hat{e}}^2(k) = \frac{\langle \delta_e(k)\delta^*_{\hat{e}}(k)\rangle^2}{\langle |\delta_e(k)|^2\rangle\langle|\delta_{\hat{e}}(k)|^2\rangle} .
\end{equation}
We observe that the noise is inversely proportional to the integrated $r^2_{e\hat{e}}$, which justifies the motivation to find the estimator with the highest possible $r^2_{\hat{e}e}$. Even for training simulations with incorrect baryonic feedback, we expect that $r_{\hat{e}e} > r_{eg}$. That is, we assume that the trained neural network is no worse at predicting electron positions in reality than the original observed galaxy field. Further, the signal of both estimators is the same, since they were properly normalized:
\begin{align}
    \left< \hat v_r(\mathbf{k}_L) \hat v^*_r(\mathbf{k}'_L) \right>_{\mathrm{signal}} = (2\pi)^3 \delta^{(3)}(\mathbf{k}_L - \mathbf{k}'_L)\, P_{v_r}(k_L).
\end{align}
Thus the lower noise leads to an increase in the signal-to-noise from the neural network method with respect to the traditional QE (as pointed out previously in \cite{Kvasiuk:2023nje}). See Sec. II of \cite{Deutsch:2017_2Dksz_estim1} for a more detailed discussion of signal and noise of the QE.

\subsection{Marginalizing over Baryonic Physics}

We now show that both the traditional QE and the neural network QE are equally robust to baryonic physics mismodeling. We compute the expectation value of $\hat v_r$ allowing for the possibility that the
\emph{true} cross power $P_{\hat{e}e}^{\rm true}$ differs from the
fiducial one $P_{\hat{e}e}^{\rm fid}$ used in the weights. This models baryonic uncertainty in the simulation.
Using \eqref{eq:vr_estimator}, the
expectation value of the estimator is
\begin{align}
    &\big\langle \hat v_r(\mathbf{k}_L) \big\rangle = \\ &=
    v_r(\mathbf{k}_L)\,
    N_{v_r}(k_L)\,\frac{K_*^2}{\chi_*^2}
    \int \frac{d^3 k_S}{(2\pi)^3}
    \int \frac{d^2 \boldsymbol{\ell}}{(2\pi)^2}\,
    \frac{P_{\hat{e}e}^{\rm fid}(k_S)\,P_{\hat{e}e}^{\rm true}(k_S)}{P^{\rm tot}_{\hat{g}\hat{g}}(k_S)\,C_{\ell}^{TT,{\rm tot}}}\,
    (2\pi)^3\delta^{(3)}\!\left(\mathbf{k}_L+\mathbf{k}_S+\frac{\boldsymbol{\ell}}{\chi_*}\right).
    \label{eq:EV_intermediate}
\end{align}
Carrying out the $\boldsymbol{\ell}$ and angular integrations as in the
normalization calculation \eqref{eq:vr_norm}, we obtain
\begin{align}
    \big\langle \hat v_r(\mathbf{k}_L) \big\rangle
    &=
    v_r(\mathbf{k}_L)\,
    N_{v_r}(k_L)\,\frac{K_*^2}{\chi_*^2}
    \int \frac{k_S\,dk_S}{2\pi}\,
    \frac{P_{\hat{e}e}^{\rm fid}(k_S)\,P_{\hat{e}e}^{\rm true}(k_S)}
         {P^{\rm tot}_{\hat{e}\hat{e}}(k_S)\,C_{\ell}^{TT,{\rm tot}}}
    \Bigg|_{\ell=k_S\chi_*}.
    \label{eq:EV_1d}
\end{align}
Using the normalization \eqref{eq:vr_norm} (with $P_{\hat{e}e}^{\rm fid}$ in the
weights),
\begin{align}
    N_{v_r}^{-1}(k_L)
    &=
    \frac{K_*^2}{\chi_*^2}
    \int \frac{k_S\,dk_S}{2\pi}\,
    \frac{\big(P_{\hat{e}e}^{\rm fid}(k_S)\big)^2}
         {P^{\rm tot}_{\hat{e}\hat{e}}(k_S)\,C_{\ell}^{TT,{\rm tot}}}
    \Bigg|_{\ell=k_S\chi_*},
    \label{eq:inv_noise}
\end{align}
we can write Eq.~\eqref{eq:EV_1d} as
\begin{align}
    \big\langle \hat v_r(\mathbf{k}_L) \big\rangle
    &=
    v_r(\mathbf{k}_L)\,
    \frac{
      \displaystyle
      \int \frac{k_S\,dk_S}{2\pi}\,
      \dfrac{P_{\hat{e}e}^{\rm fid}(k_S)\,P_{\hat{e}e}^{\rm true}(k_S)}
            {P^{\rm tot}_{\hat{e}\hat{e}}(k_S)\,C_{\ell}^{TT,{\rm tot}}}
      \Big|_{\ell=k_S\chi_*}
    }{
      \displaystyle
      \int \frac{k_S\,dk_S}{2\pi}\,
      \dfrac{\big(P_{\hat{e}e}^{\rm fid}(k_S)\big)^2}
            {P^{\rm tot}_{\hat{e}\hat{e}}(k_S)\,C_{\ell}^{TT,{\rm tot}}}
      \Big|_{\ell=k_S\chi_*}
    }.
\end{align}
Defining the kernel
\begin{align}
    F(k_S) \equiv
    \frac{k_S}{2\pi}\,\frac{P_{\hat{e}e}^{\rm fid}(k_S)}
               {P^{\rm tot}_{\hat{e}\hat{e}}(k_S)\,C_{\ell}^{TT,{\rm tot}}}
    \Bigg|_{\ell=k_S\chi_*},
\end{align}
we finally obtain
\begin{align}
    \big\langle \hat v_r(\mathbf{k}_L) \big\rangle
    &= b_v\,v_r(\mathbf{k}_L),\\[4pt]
    b_v
    &=
    \frac{\displaystyle \int dk_S\,F(k_S)\,P_{\hat{e}e}^{\rm true}(k_S)}
         {\displaystyle \int dk_S\,F(k_S)\,P_{\hat{e}e}^{\rm fid}(k_S)}.
\end{align}
This defines the velocity reconstruction bias $b_v$, which quantifies the
multiplicative error induced by using a fiducial
$P_{\hat{e}e}^{\rm fid}$ when the true is $P_{\hat{e}e}^{\rm true}$ (in the case of the traditional kSZ QE, this was recently also noted in \cite{Tishue:2025cvp, McCarthy:2025brx})\footnote{This $b_v$ bias parameter is different from the one introduced in Section~(\ref{sec:robust}), which is $b(k)=P_{\hat{e}e}^{\rm fid}(k_S)/P_{ee}^{\rm true}(k_S)$, but both biases parameterize the difference between the assumed model and true distribution.}. As was mentioned, we marginalize over $b_v$ in the analysis. As we have shown above, the neural network QE will still be robust to baryonic physics in the same way as the traditional QE, since we marginalize over the associated bias parameter in both cases. 

\end{document}